\documentclass{pazhbabel_eng}

\usepackage{amsmath}
\usepackage{amsfonts}
\usepackage{amssymb}
\usepackage{amstext}
\usepackage{natbib}
\usepackage{indentfirst}
\usepackage{misccorr}

\bibpunct{(}{)}{;}{a}{}{,}

\usepackage{lscape}
\usepackage{mathtext}
\usepackage{graphicx}
\usepackage{float}

\usepackage{textcomp}
\usepackage{mathtext}

\usepackage{epsfig}
\usepackage{rotating}
\usepackage{dcolumn}
\usepackage{tabularx}

\usepackage{afterpage}
\usepackage{figsize}

\newcommand{\Mbh}{M_{\rm{BH}}}
\newcommand{\sigstar}{\sigma_\ast}
\newcommand{\Reff}{R_{\rm{eff}}}
\newcommand{\Ak}{A_{K}}
\newcommand{\mk}{m_{K}}

\newcommand{\Mkb}{M_{K,{\rm b}}}
\newcommand{\Lk}{L_{K}}
\newcommand{\Lkp}{L_{K,{\rm PSC}}}
\newcommand{\Lkb}{L_{K,\rm{b}}}
\newcommand{\Lkagn}{L_{K,\rm{AGN}}}
\newcommand{\Pagn}{P_{\rm{AGN}}}
\newcommand{\Pb}{P_{\rm{b}}}
\newcommand{\Lint}{L_{17-60~\rm{keV}}}

\newcommand{\Av}{A_{V}}
\newcommand{\Avb}{A_{V,{\rm Balm}}}
\newcommand{\Avg}{A_{V,{\rm gal}}}

\newcommand{\Mdyn}{M_{\rm{dyn}}}
\newcommand{\Mrev}{M_{\rm{rev}}}
\newcommand{\Mhb}{M_{\rm{H}\beta}}
\newcommand{\Mbulge}{M_{\rm{bulge}}}

\newcommand{\Ledd}{L_{\rm{Edd}}}
\newcommand{\Leddbulge}{L_{\rm{Edd, bulge}}}
\newcommand{\Leddbest}{L_{\rm{Edd, best}}}
\newcommand{\Lbol}{L_{\rm{bol}}}
\newcommand{\Lmir}{L_{15\,\mu\rm{m}}}

\journalinfo{2012}{38}{8}{475}[491]
\begin{document}

\title{Masses and Accretion Rates of Supermassive Black Holes in Active
Galactic Nuclei from the INTEGRAL Survey}
\author{
G.~A.~Khorunzhev \email{horge@iki.rssi.ru}, 
S.~Yu.~Sazonov , 
R.~A.~Burenin and A.~Yu.~Tkachenko 
\addresstext{Space Research Institute, Russian Academy of Sciences, Profsoyuznaya ul. 84/32, Moscow, 117997 Russia}
}
\shortauthor{}
\shorttitle{}
\submitted{}

\begin{abstract}
\textbf{Abstract}---The masses of 68 supermassive black holes (SMBHs) in nearby ($z<0.15$) active galactic
nuclei (AGNs) detected by the INTEGRAL observatory in the hard X-ray energy band (17--60 keV)
outside the Galactic plane ($|b| > 5^\circ$) have been estimated. Well-known relations between the SMBH mass and (1) the infrared luminosity of the stellar bulge (from 2MASS data) and (2) the characteristics of broad emission lines (from RTT-150 data) have been used. A comparison with the more accurate SMBH mass
estimates obtained by the reverberation-mapping technique and from direct dynamical measurements is
also made for several objects. The SMBH masses derived from the correlation with the bulge luminosity
turn out to be systematically higher than the estimates made by other methods. The ratio of the bolometric luminosity to the critical Eddington luminosity has been found for all AGNs. It ranges from 1 to 100\% for the overwhelming majority of objects.
\par
\DOI{10.1134/S1063773712080026}
\par
\englishkeywords{active galactic nuclei (AGNs), supermassive black holes (SMBHs).}

\end{abstract}

\section{INTRODUCTION}
\label{s:intro}

Measuring the masses of supermassive black
holes (SMBHs) in galactic nuclei is an important part
of the various studies of both the SMBHs themselves
and the interrelationship between the evolution of
galaxies and the growth of black holes in their central
regions.
\\
The most reliable known method of determining
the SMBH mass involves measuring the kinematics
of stars or gas motions in the gravitational field of
a central black hole. Unfortunately, since the angular
resolution of telescopes is finite, the applicability
of this method is so far restricted to nearby
galaxies: the dynamical SMBH mass estimates have
been obtained for about 50 galactic nuclei \citep{guletal09}, in the overwhelming majority normal
(inactive) ones.
\par
Measuring the SMBH masses in active galactic
nuclei (AGNs)~---~Seyfert galaxies and quasars~---~is of
particular interest, because an active growth of black
holes occurs in them. The reverberation-mapping
technique, applicable to type-1 AGNs (with broad
emission lines in the optical spectrum), is deemed to
be fairly reliable. It consists in: (1) measuring the
size of the broad-line region from the delay between
the line and continuum flux variations, (2) measuring
the gas velocity dispersion in this region from
the line width, and (3) applying the virial theorem
to determine the black hole mass from these two
characteristics. The main problem associated with
reverberation mapping is the necessity of performing
multiple spectrophotometric observations of AGNs
for several weeks or months. Therefore, so far the
SMBHmasses have been measured by this technique
only in several tens of Seyfert galaxies and quasars
\citep{kasetal00,petetal04}.
\par
Since the methods described above, which may
be considered direct ones, are difficult to apply, indirect
methods are also actively used to estimate the
SMBH masses. For example, in the case of type-1 AGNs, an empirical relation between the SMBH
mass and broad-line characteristics (width and luminosity)
(see, e.g., \citealt{vespet06}) has been widely used in recent years. A physical justification
for the existence of such a relation was offered
by \citet{dibai77}.In recent years, the SMBH masses
in AGNs (of both type 1 and type 2) also came to be
often estimated from the correlation with the mass
or velocity dispersion of the host stellar spheroid~---
an elliptical galaxy or the bulge of a spherical galaxy
\citep{fermer00,treetal02,marhun03,guletal09}.
\par
An increasingly active use of indirect methods to
determine the SMBH masses in AGNs raises the
question of how reliable such estimates are. In particular,
is it justifiable to use the correlation between the
stellar bulge properties and the central SMBH mass
established for normal galaxies in the case of active
galaxies? Previous studies suggested that the ratio
of the black hole mass to the bulge mass in AGNs
was systematically lower than that in normal galaxies
\citep{wandel99}. However, it was shown in a
number of succeeding papers that AGNs obeyed the
same correlation between the SMBH mass and the
velocity dispersion in the bulge ($M_{BH}$--$\sigstar$) as did normal
galaxies \citep{neletal04,onketal04,wooetal10,graetal11}, but, at the
same time, significant differences in the correlation
between the black holemass and the bulge luminosity
were revealed \citep{neletal04,benetal09}.
The interpretation of these results is complicated by
the fact that the $\Mbh$ estimates used for comparison
with the bulge characteristics were obtained mainly
by the reverberation-mapping technique and, consequently,
were determined to within the normalization
factor $f$ in the virial formula (see Eq.\ref{eq:virialmass}),
which can change from object to object by several times. In
addition, at least in the case of one type of AGN,
namely narrow-line Seyfert 1 galaxies, the fact that
the SMBHs have masses considerably lower than
those expected from the correlations with the bulge
parameters established for normal galaxies may be
considered to have been firmly established \citep{matetal01,wandel02,grumat04}.
As an illustration of the existing uncertainty, it can
be noted that the sample of 49 galaxies used by
\citep{guletal09} to construct the correlation
between the bulge and SMBH masses contains three
Seyfert galaxies with highly accurate (based on maser
emission) SMBH mass estimates, and the measured
black hole mass in one of them (the Circinus galaxy)
turns out to be approximately a factor of 30 lower than
that expected from the correlation with the stellar
velocity dispersion!
\par
In this paper, we applied various known methods
to determine the SMBH mass and the ratio of the
current accretion rate to the critical one for a representative
sample of nearby Seyfert galaxies drawn
from the INTEGRAL hard X-ray all-sky survey. This
sample has already been used and is being used for
various systematic studies of the local population of
AGNs. This work also allows us to reach independent
conclusions about the applicability of various methods
for estimating the SMBH masses in AGNs.

\section{THE SAMPLE}
\label{s:data}

\begin{table*}
\centering
\caption{Sample of AGNs from the INTEGRAL survey at $|b|>5^\circ$}
\label{tab:agnintro}

\smallskip

\scriptsize

\begin{tabular}{lcllcD{.}{.}{1}c}
\hline
\hline
\multicolumn{1}{c}{Object} &
\multicolumn{1}{c}{\No} &
\multicolumn{1}{c}{Type of} &
\multicolumn{1}{c}{Type of AGN} &
\multicolumn{1}{c}{$z$} &
\multicolumn{1}{c}{$D$} &
\multicolumn{1}{c}{$\lg\Lint$}
\\
\multicolumn{1}{c}{} &
\multicolumn{1}{c}{} &
\multicolumn{1}{c}{galaxy} &
\multicolumn{1}{c}{} &
\multicolumn{1}{c}{} &
\multicolumn{1}{c}{(Mpc)} &
\multicolumn{1}{c}{(erg/s)} 
\\
\hline
MRK348&1&S0a&S2&0.015&63.4&43.56\\ 
MCG-01-05-047&2&Sc&S2&0.017&72.8&43.02 \\ 
NGC788&3&S0a&S2&0.014&57.4&43.28 \\ 
LEDA138501&4&&S1&0.049&213.3&44.34 \\ 
MRK1040&5&Sbc&S1.5&0.017&70.7&43.47 \\ 
IGRJ02343+3229&6&Sb&S2&0.016&68.5&43.34 \\ 
NGC1068&7&Sb&S2&0.004&14.4*&43.14 \\ 
NGC1142&8&Sb$^{(1)}$&S2&0.029&123.0&43.93 \\ 
1H0323+342&9&Sb?$^{(2)}$&NLS1&0.061&266.7&44.37 \\ 
NGC1365&10&Sb&S1.8&0.005&17.9*&42.11 \\ 
3C111&11&E&S1,BLRG&0.049&210.1&44.62 \\ 
ESO033-G002&12&S0&S2&0.018&76.7&43.14 \\ 
IRAS05078+1626&13&S0?&S1.5&0.018&75.8&43.61 \\ 
ESO005-G004&14&Sb&S2&0.006&22.4*&42.18 \\ 
MRK3&15&S0&S2&0.013&57.0&43.43 \\ 
MRK6&16&S0a&S1.5&0.019&79.7&43.45 \\ 
IGRJ07563-4137&17&S0a?&S2&0.021&89.1&43.08 \\ 
ESO209-G012&18&Sa?&S1.5&0.041&174.4&43.79 \\ 
IRAS09149-6206&19&&S1&0.057&249.8&44.19 \\ 
MRK110&20&Sa$^{(1)}$&NLS1&0.035&151.5&44.21 \\ 
IGRJ09446-2636&21&&S1.5&0.142&658.2&45.31 \\ 
NGC2992&22&Sa&S2&0.008&30.5*&42.76 \\ 
MCG-5-23-16&23&S0&S2&0.009&35.7&43.18 \\ 
NGC3081&24&S0a&S2&0.008&32.5*&42.77 \\ 
ESO263-G013&25&Sb&S2&0.033&142.7&43.72 \\ 
NGC3227&26&Sa&S1.5&0.004&20.6*&42.67 \\ 
NGC3281&27&Sab&S2&0.011&45.1&42.98 \\ 
IGRJ10386-4947&28&&S1.5&0.060&262.1&44.09 \\ 
IGRJ10404-4625&29&S0&S2&0.024&101.7&43.42 \\ 
NGC3783&30&Sa&S1&0.010&38.5*&43.34 \\ 
IGRJ12026-5349&31&S0&S2&0.028&119.5&43.62 \\ 
NGC4151&32&Sab&S1.5&0.003&20.3*&43.37 \\ 
MRK50&33&S0a$^{(1)}$&S1&0.023&99.5&43.19 \\ 
NGC4388&34&Sb&S2&0.008&16.8*&42.79 \\ 
NGC4395&35&Sm&S1.8&0.001&4.6*&40.60 \\ 
NGC4507&36&Sab&S2&0.012&49.7&43.51 \\ 
NGC4593&37&Sb&S1&0.009&39.5*&43.04 \\ 
NGC4945&38&Scd&S2&0.002&3.8*&41.54 \\ 
ESO323-G077&39&S0&S1.2&0.015&63.4&43.13 \\ 
IGRJ13091+1137&40&Sa&XBONG&0.025&106.9&43.68 \\ 
IGRJ13149+4422&41&Sa&S2&0.037&157.2&43.81 \\ 
CENA&42&E&S2,NLRG&0.002&3.6*&41.94 \\ 
MCG-6-30-15&43&S0&S1.2&0.008&32.4&42.66 \\ 
MRK268&44&Sb$^{(1)}$&S2&0.040&171.8&43.79 \\ 
IC4329A&45&S0a&S1.2&0.016&67.7&43.95 \\ 
NGC5506&46&Sa&S1.9&0.006&28.7*&43.12 \\ 
IGRJ14552-5133&47&Sc&NLS1&0.016&67.7&42.89 \\ 
IGRJ14561-3738&48&Sa&S2&0.025&104.7&43.27 \\ 
IC4518A&49&&S2&0.016&66.4&43.11 \\ 
WKK6092&50&Sb&S1&0.016&65.9&42.94 \\ 
IGRJ16185-5928&51&Sc&NLS1&0.035&150.1&43.67 \\ 
ESO137-G34&52&S0a&S2&0.009&38.7&42.48 \\ 
IGRJ16482-3036&53&&S1&0.031&133.9&43.75 \\ 
NGC6221&54&Sc&S2&0.005&19.4*&41.94 \\ 
IGRJ16558-5203&55&&S1.2&0.054&234.9&44.29 \\ 
NGC6300&56&Sb&S2&0.004&14.3*&42.07 \\ 
IGRJ17418-1212&57&&S1&0.037&159.8&43.90 \\ 
3C390.3&58&E&S1,BLRG&0.056&244.4&44.65 \\ 
IGRJ18559+1535&59&&S1&0.084&372.3&44.58 \\ 
1H1934-063&60&E&NLS1&0.011&44.6&42.63 \\ 
NGC6814&61&Sbc&S1.5&0.005&22.8*&42.47 \\ 
CYGA&62&E&S2,NLRG&0.056&244.4&44.62 \\ 
IGRJ20286+2544&63&Sd&S2&0.014&60.0&43.16 \\ 
MRK509&64&S0a$^{(1)}$&S1.2&0.034&147.5&44.16 \\ 
NGC7172&65&Sa&S2&0.009&33.9*&42.92 \\ 
MR2251-178&66&E$^{(3)}$&S1&0.064&280.4&44.65 \\ 
NGC7469&67&Sa&S1.2&0.016&68.9&43.43 \\ 
MRK926&68&S0a$^{(1)}$&S1.5&0.047&203.0&44.25 \\ 
\hline
\end{tabular}
\begin{flushleft}
Note. (1) \cite{petetal07}, (2) \cite{zhoetal07}, (3) \cite{noretal86}.
The optical and ratio types of AGNs: S1--S2 --- Seyfert galaxy, 
NLS1 --- narrow-line Seyfert 1 galaxy, BLRG --- broad-line radio galaxy,
NLRG --- narrow-line radio galaxy, XBONG --- X-ray bright optically normal galaxy; $z$ is the redshift;
$D$ is the photometric distance($H_0$=72 km s$^{-1}$ Mpc$^{-1}$,
$\Omega_{\rm m}$=0.3, $\Omega_\Lambda$=0.7). For several nearby objects, we used accurate distances (marked by the asterisk) from the Nearby Galaxies Catalogue \citep{Tullyetal09,Tully88}.
$\Lint$ is the 17--60 keV (INTEGRAL/IBIS/ISGRI) AGN luminosity.
\end{flushleft}

\end{table*}

We used the catalog of AGNs \citep{krietal07,sazetal07}, detected in the hard X-ray
energy band (17--60 keV) by the IBIS/ISGRI instrument
\citep{Uberetal03} onboard the INTEGRAL
observatory \citep{winetal03} during the first three
and a half years of the mission, from October 2002
to June 2006. Here, we use a complete sample of
68 Seyfert 1 (33 objects) and 2 (35 objects) galaxies
that does not include blazars and objects located near
the Galactic plane ($|b|<5^\circ$).
All AGNs represent
the local Universe (the redshift of the most distant
object is $z=0.14$). The main advantage of the
sample is that it was drawn on the basis of hard X-ray
observations and, therefore, essentially does not
suffer from ordinary selection effects related to the
absorption of emission in AGNs or the difficulty of its
detection against the galaxy's bright background. It
may be considered to be a representative sample of
Seyfert galaxies in the sense that it properly reflects
the proportion of type-1 and type-2 AGNs in the local
Universe and covers almost the entire luminosity
function of Seyfert galaxies. This sample has been
used and is being actively used to systematically
study such properties of the local population of AGNs
as the luminosity function and the photoabsorption
column density distribution \citep{sazetal07},
the total broadband X-ray spectrum of the population
\citep{sazetal08}, the ratio of the X-ray and
infrared luminosities for AGNs (using data from the
Spitzer telescope; \citealt{sazetal12}), and the ratio
of the X-ray and narrow-line luminosities (Sazonov
et al., in preparation). In the subsequent work, it
would be interesting to use this sample to investigate
how different AGN emission characteristics depend
on the SMBH mass and accretion regime.
\par
Basic information about the objects being studied
is presented in Table~\ref{tab:agnintro}. The morphological types and
redshifts of the galaxies were taken mainly from the
HyperLeda and NED databases. We took the types
of AGNs characterizing their optical properties and
radio loudness from our previous papers \citep{sazetal07,sazetal12}, 
where the references to the original
sources of information are given. The photometric
distances to the objects were calculated from their
redshifts for the following cosmological parameters:
$H_0$=72 km s$^{-1}$ 
Mpc$^{-1}$, $\Omega_{\rm m}$=0.3, and $\Omega_\Lambda$=0.7.
For several nearby objects, we used more accurate
distances from the Nearby Galaxies Catalogue \citep{Tullyetal09,Tully88}.
In particular, we used the
distances to the objects to calculate the AGN luminosities
in the 17--60 keV energy band from the fluxes
measured (with an accuracy better than 20\%) by the
IBIS/ISGRI instrument \citep{sazetal07}.

\section{ESTIMATION OF THE SMBH MASSES FROM EMPIRICAL RELATIONS}
\label{s:calc}
We applied both well-known indirect methods for
determining the SMBH mass described in the Introduction
to our objects.

\subsection{ESTIMATION FROM THE INFRARED BULGE LUMINOSITY
(BASED ON 2MASS DATA)}
\label{s:calc2MASS}
In the last 10--15 years, the existence of a correlation
between the SMBH mass and properties of
the spheroidal component of the host galaxy, i.e.,
the bulge of a spiral galaxy or an entire elliptical
galaxy, has been firmly established. At the same
time, the question of precisely which characteristic
of the spheroid (mass, luminosity, velocity dispersion,
etc.) correlates more closely with the central
black holemass remains open, because different constructed
correlations are characterized by approximately
the same dispersion---the scatter of individual
values about the mean (see, \citealt{marhun03,graham07,guletal09}).
As has already been said, any such correlations were
firmly established only for normal (inactive) galaxies,
while their applicability for AGNs is still open to question.
\par
In this paper, we used the correlation between the
SMBH mass and the near-infrared luminosity of the
bulge/elliptical galaxy \citep{graham07}:
\begin{equation}
\lg\Mbh=-0.37(\pm 0.04)(\Mkb+24)+8.29(\pm 0.08)\, ,
\label{eq:graham}
\end{equation} 
where $\Mbh$ is the black hole mass in solar masses,
and $\Mkb$ is the extinction-corrected $K$-band absolute
magnitude of the bulge (or elliptical galaxy).
\par
The empirical relation (\ref{eq:graham}) was constructed by
Graham from 21 nearby galaxies with available dynamical
SMBH mass measurements. The scatter of
$\lg\Mbh$ in this sample about themean (including the
measurement errors) is 0.33.
\par
For our purposes, it is convenient to rewrite relation
(\ref{eq:graham}) via the bulge luminosity $\Lkb$ (using the
characteristics of the 2MASS $K_S$ filter: the central
wavelength is 2.16  $\mu m$, the bandwidth is $1.686\times 10^{13}$ Hz, and the spectral flux density corresponding to the zero magnitude is 666.7 Jy):
\begin{equation}
\lg\Mbh=0.925\lg\Lkb-31.23\, ,
\label{eq:graham_lum}
\end{equation}
where $\Lkb$ is measured in erg/s.
\par
The $\Mbh$--$\Lkb$ correlation is essentially equivalent
to the SMBH mass--bulge mass correlation,
because the infrared luminosity of galaxies is roughly
proportional to their stellar mass. We preferred to
use precisely this relation, mainly because almost for
all our objects the infrared luminosity of a galaxy
can be easily estimated from the publicly accessible
data of the Two-Micron All-Sky Survey (2MASS),
which covered 99\% of the sky in three bands:
$J$($\lambda=1.24~\mu m$ ), $H$($\lambda=1.66~\mu m$) и
$K_{S}$($\lambda=2.16~\mu m$).
\par
Most (62) objects of our sample are present either
in the Extended Source Catalog (XSC, \citealt{jaretal00}) or in the Large Galaxy Atlas (LGA,
\citealt{jaretal03}) of 2MASS, where the apparent
K magnitude ($m_K$) of the entire galaxy is given. This
allowed us to determine the total infrared luminosities
of the galaxies ($L_K$) corrected for interstellar extinction
(Table \ref{tab:2mass}).
\par
The active nucleus can contribute noticeably to
the galaxy's infrared luminosity, because part of the
emission (mainly the ultraviolet one) from the SMBH
accretion disk is absorbed and reprocessed in the
surrounding torus of gas and dust. We attempted to
estimate this contribution,
\begin{equation}
\Pagn=\frac{\Lkagn}{\Lk},
\label{eq:pagn}
\end{equation}
using the correlation between the infrared and hard
X-ray AGN luminosities. Such a correlation was
firmly established, for example, in our recent paper
\citep{sazetal12} through observations of the
same AGN sample from the INTEGRAL survey that
is used here with the Spitzer infrared telescope. It
follows from the results of this paper and a number of
published spectral energy distributions averaged over
representative samples of quasars \citep{elvetal94,sazetal04,ricetal06,shaetal11} that the K-band luminosity of the dusty torus is
\begin{equation}
\Lkagn\sim 0.1\Lint,
\label{eq:lagn}
\end{equation}
where $\Lint$ is the AGN luminosity measured in
the INTEGRAL energy band (Table \ref{tab:agnintro}). In this case,
the ratio $\Lkagn/\Lint$ keV can vary from object to
object by a factor of $\sim 2$.
\par
The values of $\Pagn$ found from Eqs. (\ref{eq:pagn}) and (\ref{eq:lagn}) are
given in Table 2. For the Seyfert galaxy NGC~1068,
the dusty torus is optically thick for Compton scattering.
Therefore, the luminosity measured by INTEGRAL
accounts for only a small fraction (of the order
of one percent; see, e.g., \citealt{sazetal12}) of the
true hard X-ray luminosity of the hot accretion disk
corona. We estimated this luminosity from Eqs. (15)
and (16) given below and only then did we calculate
$\Pagn$ from Eqs.~(\ref{eq:pagn}) and (\ref{eq:lagn}).
\par
The effective angular size of the galaxy in the
2MASS image ($\Reff$), the radius within which half
of the galaxy's total K-band luminosity is contained,
can serve as an additional indicator of the contribution
from the active nucleus to the galaxy's infrared luminosity.
The values of $\Reff$ from the XSC and LGA are
listed in Table~2. They should be compared with the
2MASS angular resolution, which is $\sim$ 2".5 \citep{skretal06}: if $\Reff$ is only a few arcsec, then this may
indicate that the contribution from the active nucleus
to $\Lk$ is dominant. However, in this case, it should be
kept in mind that the angular size of the galaxy can be
small per se for the most distant objects of our sample.
\par
For 17 objects, the contribution from the active
nucleus to the galaxy's infrared luminosity estimated
from the X-ray luminosity turned out to be greater
than 50\% ($\Pagn>0.5$). Therefore, (given some uncertainty
associated with Eq. (\ref{eq:lagn})) it turns out to be
impossible to reliably subtract the AGN contribution
from the galaxy's total luminosity for such objects,
and the values of $\Lk$ measured for them can be
used only to obtain upper limits on the SMBH mass
from its correlation with the bulge luminosity. In
addition, in all these cases, except for MRK~348
and CYG~A, the galaxy's size $\Reff\lesssim 5$'', which also
indirectly points to a substantial contribution from the
central dusty torus to $\Lk$.
\par
The next step in our calculations is to estimate
the bulge fraction ($\Pb$) in the galaxy's K-band luminosity
(i.e., basically the bulge mass fraction). Unfortunately,
an accurate allowance cannot be made
for most of the sample objects, because many of the
galaxies are at a distance of the order of or greater
than 100 Mpc and there are no optical or infrared
(2MASS) images for them with an angular resolution
high enough for the nucleus, bulge, and disk emission
components to be reliably separated. Therefore, we
decided to resort to a less accurate method, more
specifically, to assign some mean expected values
of $\Pb$ to the galaxies based on their morphological
types listed in Table~\ref{tab:agnintro}. This approach is justified,
because the bulge mass fraction correlates with the
type of galaxy, although there exists an appreciable
scatter of $\Pb$  values among galaxies of the same type.
There is no uncertainty only for elliptical galaxies, for
which, obviously, $\Pb$ = 1. For the remaining types,
we adopted the following values based on the results
from \citealt{lauetal07,grawor08,lauetal10}: $\Pb=0.25$ (S0--Sab), 0.2 (Sb), 0.13 (Sbc), 0.08 (Sc), 0.06 (Scd and later types). For four galaxies (IGR J09446-2636, IGR J10386-4947, IGR J16482-3036, IGR J18559+1535) whose
morphological types were not established but which,
nevertheless, are present in the 2MASS XSC, we
adopted $\Pb=0.25$ as a rough estimate of the bulge
fraction.
\par
Thus, we calculated the K-band bulge luminosity for most of the objects as
\begin{equation}
\Lkb=\Pb(1-\Pagn)\Lk,
\label{eq:lk_lkb}
\end{equation}
and obtained the following upper limits for the objects
in which the AGN contribution is dominant ($\Pagn>~0.5$):
\begin{equation}
\Lkb<\Pb\Lk.
\end{equation}
\par
Five objects (IRAS 05078+1626, IRAS 09149-6206, IC 4518A, IGR J16558-5203, 
IGR J17418-1212) enter neither into the Extended Source Catalog
nor into the Large Galaxy Atlas, but they are
contained only in the 2MASS Point Source Catalog
(PSC, \citealt{skretal06}). Since the method of
calculation described above cannot be used for these
objects, we made rougher estimates of the bulge luminosity
for them as follows. The PSC provides the
apparent $K$ magnitudes of the sources measured in
an aperture with a radius of 4". This angular size
corresponds to a linear size from $\sim 1.3$ to $\sim 5$ kpc for
the galaxies listed above, i.e., it roughly corresponds
to the size of the bulges in typical galaxies. Therefore,
we assumed that the measured magnitude of the point
infrared source roughly corresponded to the luminosity
of the galactic bulge, from which, however, the
expected AGN fraction should be subtracted. Thus,
we used modified formulas for these five objects:
\begin{equation}
\Pagn=\frac{\Lkagn}{\Lkp},
\end{equation}
\begin{equation}
\Lkb=(1-\Pagn)\Lkp,
\end{equation}
where $\Lkp$ is the luminosity of the PSC point source.
\par
Since the Seyfert galaxy LEDA 138501 is close
to the Galactic plane in the sky, there are no reliable
2MASS photometric data for it. In addition, the
morphological type of the galaxy is unknown. As a
result, this is the only object for which even a reliable
upper limit on the SMBH mass cannot be established
from its correlation with the stellar bulge luminosity.
The SMBH mass estimates (or upper limits) for the
remaining 67 objects are presented in Table ~\ref{tab:2mass}.
\par
Apart from the $\Mbh$ estimates themselves, their
uncertainties are also important. Since all of the
objects being studied are bright 2MASS sources, the
errors related to the photometric measurements of
their infrared luminosities may be neglected. The
main uncertainty in our calculations is associated
with the determination of the bulge fraction in the
galaxy's infrared luminosity. Based on the results
of the statistical studies of galaxies mentioned above
(see, e.g., \citealt{lauetal10}), we can estimate
the systematic error in $\lg\Pb$ and, consequently,
$\lg\Lkb$ from Eq.~(\ref{eq:lk_lkb}) for galaxies with known morphological
types to be $\sim 0.3$, i.e., the bulge luminosity
can be determined to within a factor of 2. For galaxies
with unknown morphologies, the error in $\lg\Lkb$ is,
obviously, $\sim 0.5$---this corresponds to a scatter of $\Pb$
values from $\sim 0.08$ to $\sim 0.8$ about the presumed value
of 0.25, i.e., almost any type of galaxy is admitted. The
same large error ($\sim 0.5$) can also be assigned to the
objects for which the bulge luminosity was estimated
from the luminosity of the PSC point infrared source.
Finally, we can assign an uncertainty of $\sim 0.1$ to the
values of $\lg\Lkb$ obtained for elliptical galaxies to
take into account the photometric measurement errors
in 2MASS. Because of the approximate proportionality
of relation (\ref{eq:graham_lum}) between $\Lkb$ and $\Mbh$, the
ultimate errors in $\lg\Mbh$ found from the correlation
with the bulge luminosity are expected to be the same
as the errors in $\lg\Lkb$ given above. These uncertainties
are taken into account in the subsequent
statistical analysis. 
\par
Of course, the considerable uncertainty can also
be associated with the very use of the $\Mbh$--$\Lkb$ correlation
to determine the SMBH masses in the AGNs
of our sample, but we can attempt to estimate its
value only by comparing the mass estimates obtained
by this method with those obtained by other methods
(see below).

\begin{table*}
\centering
\caption{Estimates of the SMBH masses from their correlation with the bulge luminosity}
\label{tab:2mass}
\smallskip

\scriptsize
\begin{tabular}{lcD{.}{.}{1}D{.}{.}{2}D{.}{.}{2}D{.}{.}{2}lD{.}{.}{2}D{.}{.}{2}D{.}{.}{2}}
\hline
\hline
\multicolumn{1}{c}{Object} &
\multicolumn{1}{c}{\No} &
\multicolumn{1}{c}{$\Reff$} &
\multicolumn{1}{c}{$\mk$} &
\multicolumn{1}{c}{$\Ak$} &
\multicolumn{1}{c}{$\lg\Lk$} &
\multicolumn{1}{c}{$\Pagn$} &
\multicolumn{1}{c}{$\Pb$} &
\multicolumn{1}{c}{$\lg\Lkb$} &
\multicolumn{1}{c}{$\lg\Mbh$} 
\\   
\multicolumn{1}{c}{} &   
\multicolumn{1}{c}{} &  
\multicolumn{1}{c}{(arcsec)} &  
\multicolumn{1}{c}{} &
\multicolumn{1}{c}{} & 
\multicolumn{1}{c}{(erg/s)} & 
\multicolumn{1}{c}{} &
\multicolumn{1}{c}{} &
\multicolumn{1}{c}{(erg/s)} &
\multicolumn{1}{c}{($M_\odot)$}    
\\
\hline

MRK348&1&9.0&10.10&0.024&42.70&0.72&0.25&<42.10&<7.71\\ 
MCG-01-05-047&2&23.9&9.39&0.010&43.10&0.08&0.08&41.97&7.58\\ 
NGC788&3&14.0&9.07&0.010&43.02&0.18&0.25&42.33&7.92\\ 
LEDA138501&4&&&&&&&&\\ 
MRK1040&5&17.8&9.27&0.035&43.13&0.22&0.13&42.14&7.75\\ 
IGRJ02343+3229&6&24.1&8.77&0.036&43.31&0.11&0.2&42.56&8.13\\ 
NGC1068&7&15.2&5.79&0.012&43.13&0.30&0.2&42.28&7.88\\ 
NGC1142&8&27.0&9.02&0.026&43.71&0.16&0.2&42.94&8.48\\ 
1H0323+342&9&2.4&11.79&0.078&43.30&1.19&0.2&<42.60&<8.17\\ 
NGC1365&10&59.5&6.37&0.007&43.09&0.01&0.2&42.38&7.97\\ 
3C111&11&3.1&11.38&0.605&43.46&1.44&1&<43.46&<8.97\\ 
ESO033-G002&12&6.0&10.03&0.053&42.91&0.17&0.25&42.23&7.82\\ 
IRAS05078+1626&13&&*11.61&0.110&42.29&2.10&&\lesssim42.29&\lesssim7.88\\ 
ESO005-G004&14&27.5&8.14&0.052&42.59&0.04&0.2&41.88&7.50\\ 
MRK3&15&8.6&8.97&0.069&43.08&0.22&0.25&42.37&7.96\\ 
MRK6&16&2.8&9.56&0.050&43.13&0.21&0.25&42.42&8.01\\ 
IGRJ07563-4137&17&5.2&10.13&0.283&43.09&0.10&0.25&42.44&8.02\\ 
ESO209-G012&18&4.2&10.09&0.095&43.62&0.15&0.25&42.94&8.49\\ 
IRAS09149-6206&19&&*9.43&0.067&44.18&0.10&&\backsimeq44.13&\backsimeq9.59\\ 
MRK110&20&2.2&11.80&0.005&42.77&2.74&0.25&<42.17&<7.77\\ 
IGRJ09446-2636&21&2.5&12.71&0.031&43.69&4.14&*0.25&\lesssim43.09&\lesssim8.62\\ 
NGC2992&22&14.2&8.60&0.022&42.67&0.12&0.25&42.01&7.62\\ 
MCG-5-23-16&23&5.2&9.35&0.040&42.51&0.47&0.25&41.63&7.28\\ 
NGC3081&24&21.6&8.91&0.020&42.60&0.15&0.25&41.92&7.55\\ 
ESO263-G013&25&17.9&10.53&0.062&43.25&0.30&0.2&42.40&7.98\\ 
NGC3227&26&38.7&7.64&0.008&42.70&0.09&0.25&42.06&7.67\\ 
NGC3281&27&32.6&8.31&0.035&43.13&0.07&0.25&42.49&8.07\\ 
IGRJ10386-4947&28&2.2&11.82&0.179&43.31&0.60&*0.25&\lesssim42.71&\lesssim8.27\\ 
IGRJ10404-4625&29&5.4&10.38&0.058&43.01&0.25&0.25&42.28&7.88\\ 
NGC3783&30&11.9&8.65&0.044&42.86&0.30&0.25&42.10&7.71\\ 
IGRJ12026-5349&31&2.6&9.95&0.075&43.33&0.19&0.25&42.64&8.20\\ 
NGC4151&32&16.6&7.38&0.010&42.80&0.38&0.25&41.99&7.61\\ 
MRK50&33&3.5&12.00&0.006&42.33&0.73&0.25&<41.73&<7.36\\ 
NGC4388&34&36.1&8.00&0.012&42.38&0.26&0.2&41.56&7.20\\ 
NGC4395&35&48.5&9.98&0.006&40.47&0.14&0.06&39.18&5.01\\ 
NGC4507&36&11.9&8.87&0.036&42.99&0.33&0.25&42.21&7.81\\ 
NGC4593&37&30.2&7.99&0.009&43.13&0.08&0.2&42.40&7.98\\ 
NGC4945&38&172.7&4.48&0.065&42.52&0.01&0.06&41.30&6.96\\ 
ESO323-G077&39&3.0&8.80&0.037&43.23&0.08&0.25&42.59&8.16\\ 
IGRJ13091+1137&40&9.9&10.18&0.009&43.12&0.36&0.25&42.32&7.91\\ 
IGRJ13149+4422&41&6.0&10.82&0.007&43.20&0.41&0.25&42.36&7.95\\ 
CENA&42&82.6&3.94&0.042&42.68&0.02&1&42.67&8.24\\ 
MCG-6-30-15&43&3.2&9.58&0.023&42.33&0.22&0.25&41.62&7.26\\ 
MRK268&44&4.5&10.88&0.006&43.25&0.35&0.2&42.37&7.95\\ 
IC4329A&45&2.7&8.80&0.022&43.28&0.47&0.25&42.40&7.98\\ 
NGC5506&46&12.4&8.19&0.022&42.78&0.22&0.25&42.07&7.68\\ 
IGRJ14552-5133&47&6.7&10.36&0.243&42.74&0.14&0.08&41.58&7.23\\ 
IGRJ14561-3738&48&9.6&9.63&0.031&43.33&0.09&0.25&42.69&8.25\\ 
IC4518A&49&&*11.73&0.058&42.10&1.01&&\lesssim42.10&\lesssim7.71\\ 
WKK6092&50&12.1&10.24&0.071&42.70&0.17&0.2&41.92&7.54\\ 
IGRJ16185-5928&51&4.0&11.14&0.118&43.07&0.40&0.08&41.76&7.39\\ 
ESO137-G34&52&23.7&8.26&0.123&43.05&0.03&0.25&42.44&8.02\\ 
IGRJ16482-3036&53&3.0&11.34&0.124&42.90&0.71&*0.25&\lesssim42.30&\lesssim7.89\\ 
NGC6221&54&42.4&7.12&0.061&42.88&0.01&0.08&41.78&7.41\\ 
IGRJ16558-5203&55&&*10.88&0.142&43.58&0.52&&\lesssim43.58&\lesssim9.07\\ 
NGC6300&56&54.3&6.93&0.036&42.68&0.02&0.2&41.97&7.59\\ 
IGRJ17418-1212&57&&*11.29&0.210&43.10&0.62&&\lesssim43.10&\lesssim8.64\\ 
3C390.3&58&2.1&11.54&0.026&43.30&2.24&1&<43.30&<8.82\\ 
IGRJ18559+1535&59&4.3&12.53&0.346&43.40&1.53&*0.25&\lesssim42.79&\lesssim8.35\\ 
1H1934-063&60&5.3&9.67&0.108&42.60&0.11&1&42.56&8.13\\ 
NGC6814&61&31.8&7.66&0.067&42.81&0.05&0.13&41.90&7.53\\ 
CYGA&62&13.4&10.28&0.140&43.85&0.59&1&<43.85&<9.33\\ 
IGRJ20286+2544&63&8.1&9.87&0.164&42.80&0.23&0.06&41.47&7.12\\ 
MRK509&64&2.1&10.01&0.021&43.47&0.49&0.25&42.58&8.15\\ 
NGC7172&65&19.0&8.32&0.010&42.87&0.11&0.25&42.21&7.81\\ 
MR2251-178&66&1.9&11.13&0.014&43.58&1.18&1&<43.58&<9.08\\ 
NGC7469&67&4.7&8.85&0.025&43.28&0.14&0.25&42.61&8.18\\ 
MRK926&68&5.4&10.64&0.015&43.49&0.57&0.25&<42.89&<8.44\\ 

\hline
\end{tabular}
\begin{flushleft}
Note. $\Reff$ is the effective angular size of the galaxy (from the XSC and LGA of 2MASS); $\mk$ is the apparent magnitude of the entire galaxy (from the XSC and LGA), or (marked by the asterisk) the point source (from the PSC of 2MASS); $\Ak$ is the interstellar extinction toward the source \citep{schetal98};
$\Lk$ is the extinction-corrected luminosity of the entire galaxy or the PSC point
source; $\Pagn$ is the AGN fraction in $\Lk$; 
$\Pb$ is the bulge mass fraction taken to be 0.25 in the absence of morphological information
about the galaxy (marked by the asterisk); $\Lkb$ is the bulge luminosity; $\Mbh$ is the SMBH mass.
\end{flushleft}

\end{table*}

\subsection{ESTIMATION FROM BROAD-LINE PARAMETERS (BASED ON RTT-150 OBSERVATIONS)}
\label{s:calcRTT}
The broad emission lines present in the spectra of
type-1 AGNs originate in the clouds of photoionized
gas located inside the sphere of gravitational influence
of the SMBH. This provides the fundamental possibility
of using the virial theorem to determine the
SMBH mass:
\begin{equation}
\Mbh= f\frac{Rv^2}{G}\, ,
\label{eq:virialmass}
\end{equation}
where $R$ is the characteristic distance from the
SMBH at which the emission in a particular broad
line is generated, $v$ is the gas velocity dispersion in
this region, $G$ is the gravitational constant, and $f$ is a
coefficient dependent on the structure and orientation
of the broad-line region.
\par
The characteristic velocity v is easy to estimate
from the Doppler broadening of spectral lines. As
regards the size of the region $R$, it can be measured
by the reverberation-mapping technique. However,
this suggests a spectroscopic monitoring of AGNs
for many nights, which is by no means always possible.
Therefore, in recent years, an indirect method
based on the estimation of the size of the broadline
region from the AGN luminosity ($L$) either in
the broadest line or in the continuum has been actively
used. Dibai \citeyearpar{dibai77} predicted a power law
$R\propto L^\alpha$ with a slope $\alpha=1/3$. It follows from reverberation measurements
that $\alpha\sim 0.5$--0.7 \citep{wanetal99,kasetal00,kasetal05,vespet06}. The fundamental advantage of this method is that the SMBH mass in AGNs can be
estimated from a single measurement of the line width
and the line or continuum luminosity.
\par
We applied the empirical relation between $\Mbh$
and parameters of the broad Balmer H$_\beta$ line ($\lambda=4861$~\AA) from \citep{vespet06} to
estimate the SMBH masses in the type-1 AGNs from our sample:

\begin{multline}
\lg\Mbh=  \lg\left[\left(\frac{{\rm FWHM} ({\rm H}_\beta)}{1000~\mbox{km
      s}^{-1}}\right)^2 \right. \times \\
      \times \left. \left(\frac{L({\rm H}_\beta)}{10^{42}~\mbox{erg
      s}^{-1}}\right)^{0.63}\right]
     +6.67\, , 
\label{eq:vespet}
\end{multline}
where the black hole mass $\Mbh$ is in solar masses,
$FWHM$ is the full width at half maximum of the line,
and $L$ is the line luminosity corrected for the absorption
on the line of sight. Relation (\ref{eq:vespet}) is characterized
by an intrinsic scatter (of $\lg\Mbh$ values about the
mean) of $\sim 0.43$ \citep{vespet06}.
\par
It is important to note that the virial formula (\ref{eq:virialmass})
is defined to within the scale factor $f$, because there
exists a fundamental uncertainty in how the gas velocity
field is structured in the broad-line regions. For
example, $f=3$ for a spherically symmetric broad-line
region \citep{wanetal99,kasetal00} and
$f\to\infty$ for a Keplerian disk whose plane is oriented
toward the observer. Therefore, in several papers,
the mean $\langle f\rangle$ for the sample was found in such a
way that the SMBH masses measured by the reverberation
technique satisfied the $\Mbh$---$\sigstar$ correlation
for normal galaxies. Onken et al. (2004) obtained
$\langle f\rangle=5.5$ precisely in this way. This value was then
used by \citealt{vespet06} to derive
the empirical formula (\ref{eq:vespet}). Thus, this formula implies
that the standard $\Mbh$---$\sigstar$ relation must hold for
AGNs.
\par
In 2008--2010, a series of spectroscopic observations
was performed with the Russian-Turkish 1.5-m telescope (RTT-150) for a sample of 19 Seyfert\,1 galaxies from the INTEGRAL survey with declinations ${\delta>-20^\circ}$. The observations were carried out with the TFOSC low- and medium-resolution spectrometer. At the first stage of these observations, broadband (3230--9120 \AA) spectra were taken for all
objects with a resolution of 12 \AA\  using grism \#15 and
a 67-$\mu m$ (1.74-arcsec) slit. Subsequently, higher resolution
spectra were taken for some of the objects
in the 3900--6800~\AA\  band with a resolution of 4~\AA\
(grism \#7, a 54 $\mu m$ slit, 1.40 arcsec) and the 5850--8270~\AA\ band with a resolution of 3~\AA\ (grism \#8, a 54-$\mu m$ slit, 1.40 arcsec).
\par
We reduced the optical observations and calibrated
the spectra using the \emph{IRAF}\footnote{http://iraf.noao.edu} standard software. The
bias image was subtracted from the two-dimensional
spectra; subsequently, the images were aligned using
the flat-field spectra obtained from the spectrum of
a halogen lamp. The one-dimensional spectra of
the objects were extracted in a linear aperture that
was centered on the object's emission maximum and
had a size large enough for only a negligible fraction
of the flux in broad Balmer lines to fall outside this
aperture. The spectral density of the emission was
calibrated in a standard way using observations of
spectrophotometric standard stars.
\par
In most cases, because of the atmospheric jitter,
the angular resolution during our observations was
such that the size of a point source exceeded the slit
size. Therefore, we corrected the fluxes in all of our
spectra for the slit by assuming the slit to cut off
part of the flux from a source in the shape of a two-dimensional
Gaussian.
\par
The subsequent scientific analysis consisted in
subtracting the continuum described by a highdegree
polynomial and modeling the line spectra
near H$_\alpha$ и H$_{\beta}$. The model of the spectrum
near H$_{\beta}$ consisted of broad and narrow allowed H$_{\beta}$
lines ($\lambda=4861.3$ \AA) and two forbidden [OIII] lines
($\lambda_1=4958.9$, $\lambda_2=5006.8$~\AA). The ratio of the [OIII]
doublet line fluxes was fixed: $F_2/F_1=2.91$. The
profiles of all lines were assumed to be Gaussian.
The intensity, broadening (measured in km/s), and
systemic velocity of all the listed lines were free model
parameters, but the broadenings and velocities of all
narrow (H$_\beta$ и [OIII]) lines were assumed to be
identical (which is reasonable, because these lines
originate approximately in the same gas). For the
spectrum near H$_{\alpha}$, we used a similar model that, in
addition to the broad and narrow H$_{\alpha}$ lines, included
the doublet of forbidden [NII] lines ($\lambda_1=6548.0$, $\lambda_2=6583.4$ \AA, $F_2/F_1=2.96$). The broadenings and
velocities of the narrow H$_{\alpha}$ and [NII] lines were
assumed to be identical. All of the measured line
broadenings were then corrected for the spectral
resolution of the measurements.

\begin{figure}
\includegraphics[bb=10 160 550 670,width=\columnwidth]{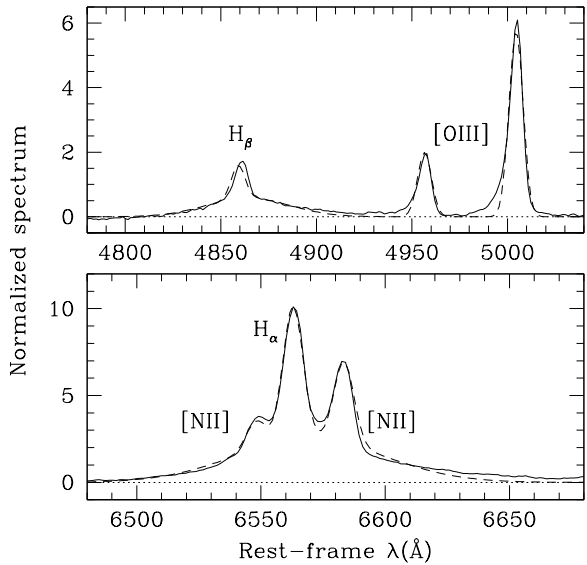}
\caption{Example of modeling the RTT-150 spectrum. The solid curve indicates the spectrum of the Seyfert galaxy NGC~7469 near H$_\beta$ and H$_\alpha$ taken with a resolution $\sim 4$~\AA. The model (dashed curve) includes the broad and narrow Balmer lines as well as the narrow forbidden [OIII]
 and [NII]. The continuum was described by a
polynomial and was subtracted.} 
\label{fig:spectra}
\end{figure}

\par
The described model allowed the parameters of the
lines of interest to us to be measured fairly reliably
for most of the objects (Table~\ref{tab:RTT}). For the narrow-line
Seyfert 1 galaxy 1H~0323+342, we used a simplified
model in which the H$_\alpha$ and H$_{\beta}$ lines had only broad
components (nevertheless, the complete model was
used for two other AGNs of the same type, MRK~110
and 1H~1934-063), because the narrow components
of the Balmer lines could not be distinguished. An
example of modeling the spectrum is shown in Fig.~{\ref{fig:spectra}}.
The broad lines in the spectra of two objects, MRK~6
and 3C~390.3, have a complex shape and are poorly
described by a Gaussian. Therefore, the parameters
measured for them cannot be considered reliable and
were excluded from the subsequent analysis.
\par
\begin{table*}
\tabcolsep=1.5mm

\caption{Estimates of the SMBH masses from Balmer-line parameters}
\label{tab:RTT}

\smallskip

\scriptsize

\begin{tabular}{lcr@{$\pm$}lr@{$\pm$}lccr@{$\pm$}lc@{$\pm$}cr@{$\pm$}lc@{$\pm$}cc@{$\pm$}c}

\hline
\hline
\multicolumn{1}{c}{Объект} &
\multicolumn{1}{c}{\No} &
\multicolumn{2}{c}{$F({\rm H}_\alpha)$} &
\multicolumn{2}{c}{$F({\rm H}_\beta)$} &
\multicolumn{1}{c}{$\Avb$} &
\multicolumn{1}{c}{$\Avg$} &
\multicolumn{2}{c}{FWHM(H$_\alpha$)} &
\multicolumn{2}{c}{$\lg L$(H$_\alpha$)} &
\multicolumn{2}{c}{FWHM(H$_\beta$)} &
\multicolumn{2}{c}{$\lg L$(H$_\beta$)} &
\multicolumn{2}{c}{$\lg\Mbh$}

\\   
\multicolumn{1}{c}{} &
\multicolumn{1}{c}{} &
\multicolumn{4}{c}{($10^{-13}\ \mbox{erg}~\mbox{s}^{-1}~\mbox{cm}^{-2}$)} &
\multicolumn{1}{c}{} &
\multicolumn{1}{c}{} &
\multicolumn{2}{c}{(km/s)} &
\multicolumn{2}{c}{(erg/s)} &
\multicolumn{2}{c}{(km/s)} &
\multicolumn{2}{c}{(erg/s)} &
\multicolumn{2}{c}{($M_\odot$)} 
\\
\hline

LEDA138501&4& 8.22&0.20& 3.14&0.30&0&\textbf{0.53}&5275&144&42.82&0.01&5981&530&42.48&0.04&8.53&0.08\\ 
MRK1040&5& 6.13&0.17& 1.07&0.11&\textbf{1.96$\pm$0.34}&0.32&3125&80&42.21&0.11&4349&497&41.72&0.17&7.77&0.14\\ 
1H0323+342&9& 1.18&0.17& 0.52&0.05&0&\textbf{0.71}&1879&99&42.23&0.06&1634&181&41.98&0.04&7.08&0.10\\ 
3C111&11& 1.73&0.03& 0.12&0.01&4.85$\pm$0.32&\textbf{5.46}&4757&89&43.75&0.01&6021&608&43.34&0.04&9.08&0.09\\ 
IRAS05078+1626&13& 3.33&0.10& 0.61&0.08&\textbf{1.81$\pm$0.41}&0.99&4034&132&41.95&0.13&4549&608&41.47&0.20&7.65&0.17\\ 
MRK110&20& 7.38&0.59& 2.21&0.10&0.27$\pm$0.29&\textbf{0.04}&2546&99&42.32&0.10&2369&141&41.80&0.14&7.29&0.10\\ 
NGC3227&26&24.10&1.55& 6.52&0.86&\textbf{0.59$\pm$0.46}&0.08&3504&200&41.28&0.15&3704&584&40.80&0.22&7.05&0.20\\ 
NGC4151&32&75.80&1.98&20.90&1.50&\textbf{0.53$\pm$0.24}&0.09&5030&160&41.75&0.08&5430&469&41.26&0.12&7.67&0.10\\ 
MRK50&33& 2.47&0.11& 0.86&0.09&0&\textbf{0.05}&4064&212&41.48&0.02&4192&495&41.03&0.05&7.30&0.11\\ 
NGC4593&37& 6.74&0.33& 2.34&0.26&0&\textbf{0.08}&3645&200&41.13&0.02&4260&563&40.68&0.05&7.10&0.12\\ 
IGRJ17418-1212&57& 1.19&0.10& 0.20&0.07&2.04$\pm$1.07&\textbf{1.90}&6137&469&42.18&0.35&5586&1733&41.68&0.52&7.96&0.42\\ 
1H1934-063&60& 3.80&0.33& 1.00&0.14&0.68$\pm$0.51&\textbf{0.97}&1768&64&41.27&0.04&1418&200&40.83&0.06&6.24&0.13\\ 
NGC6814&61& 6.12&0.31& 2.04&0.25&0&\textbf{0.79}&2998&146&40.84&0.02&4015&589&40.47&0.05&6.91&0.13\\ 
MRK509&64&24.00&0.81& 7.23&0.47&0.26$\pm$0.23&\textbf{0.19}&4024&127&42.86&0.08&4580&372&42.36&0.11&8.22&0.10\\ 
MR2251-178&66&10.80&0.42& 3.75&0.49&0&\textbf{0.13}&5564&243&43.05&0.02&6735&791&42.61&0.06&8.71&0.11\\ 
NGC7469&67&22.20&0.83& 5.89&0.54&\textbf{0.65$\pm$0.31}&0.23&3205&118&42.31&0.10&3148&346&41.83&0.15&7.56&0.13\\ 
MRK926&68& 8.07&0.23& 3.00&0.28&0&\textbf{0.14}&6709&250&42.65&0.01&9012&1050&42.24&0.04&8.73&0.10\\     

\hline
\end{tabular}
Note. $F({\rm H}_\alpha, {\rm H}_\beta)$ are the extinction-uncorrected line fluxes;
$\Avb$ is the extinction from the Balmer-line ratio;
$\Avg$ is the extinction in the Galaxy \citep{schetal98}.
FWHM (H$_\alpha$, H$_\beta$) are the full widths at half maximum of the lines;
$L({\rm H}_\alpha, {\rm H}_\beta)$ are the
extinction-corrected line luminosities (we used the boldfaced value of $\Avg$ или $\Avb$); 
$\Mbh$ is the SMBH mass.

\end{table*}

The extinction in the path from the broad-line
region to the observer was estimated from the ratio
of the Balmer-line fluxes (Balmer decrement):
\begin{equation}
\Avb=7.21\lg\frac{F({\rm H}_\alpha)}{B_0F({\rm H}_\beta)},
\label{eq:extinct}
\end{equation}
where $B_0=F_0({\rm H}_\alpha)/F_0({\rm H}_\beta)$ is the expected Balmer
decrement for broad lines in AGNs without extinction,
which was assumed to be 3.06 \citep{donetal08}.
\par
The value of $\Avb$ measured in this way was
then compared with the interstellar extinction in our
Galaxy toward the object $\Avg$ \citep{schetal98}.
If $\Avb$ turned out to exceed $\Avg$ by more than
$1\sigma$ (the error of our measurement), then we used
$\Av=\Avb$ for the subsequent calculations, i.e.,
we assumed the broad-line emission to be absorbed
not only in the interstellar medium of our Galaxy but
also in the object itself (in the galaxy or its nucleus).
Otherwise, we used $\Av=\Avg$. The final correction
of the measured line fluxes for extinction was made by
a standard method \citep{caretal89}: $A({\rm H}_\alpha)=0.8177\Av$,  $A({\rm H}_\beta)=1.1643\Av$. The derived line luminosities
are given in Table ~\ref{tab:RTT}.
\par
The last column in Table~\ref{tab:RTT} gives the SMBH
masses calculated from Eq. (\ref{eq:vespet}). The presented
errors take into account the uncertainties related to
both the measurement of the H$_\beta$ parameters and the
extinction correction. The latter dominates in the
objects in which an additional (with respect to the
extinction in the Galaxy) extinction was revealed by
the Balmer decrement. The uncertainty associated
with Eq.~(\ref{eq:vespet}) itself was disregarded.

\begin{figure}[!]
\centering
\SetFigLayout[3]{3}{1}
\subfigure{
\includegraphics[width=0.8\columnwidth,viewport= 0 0 500 500,clip]{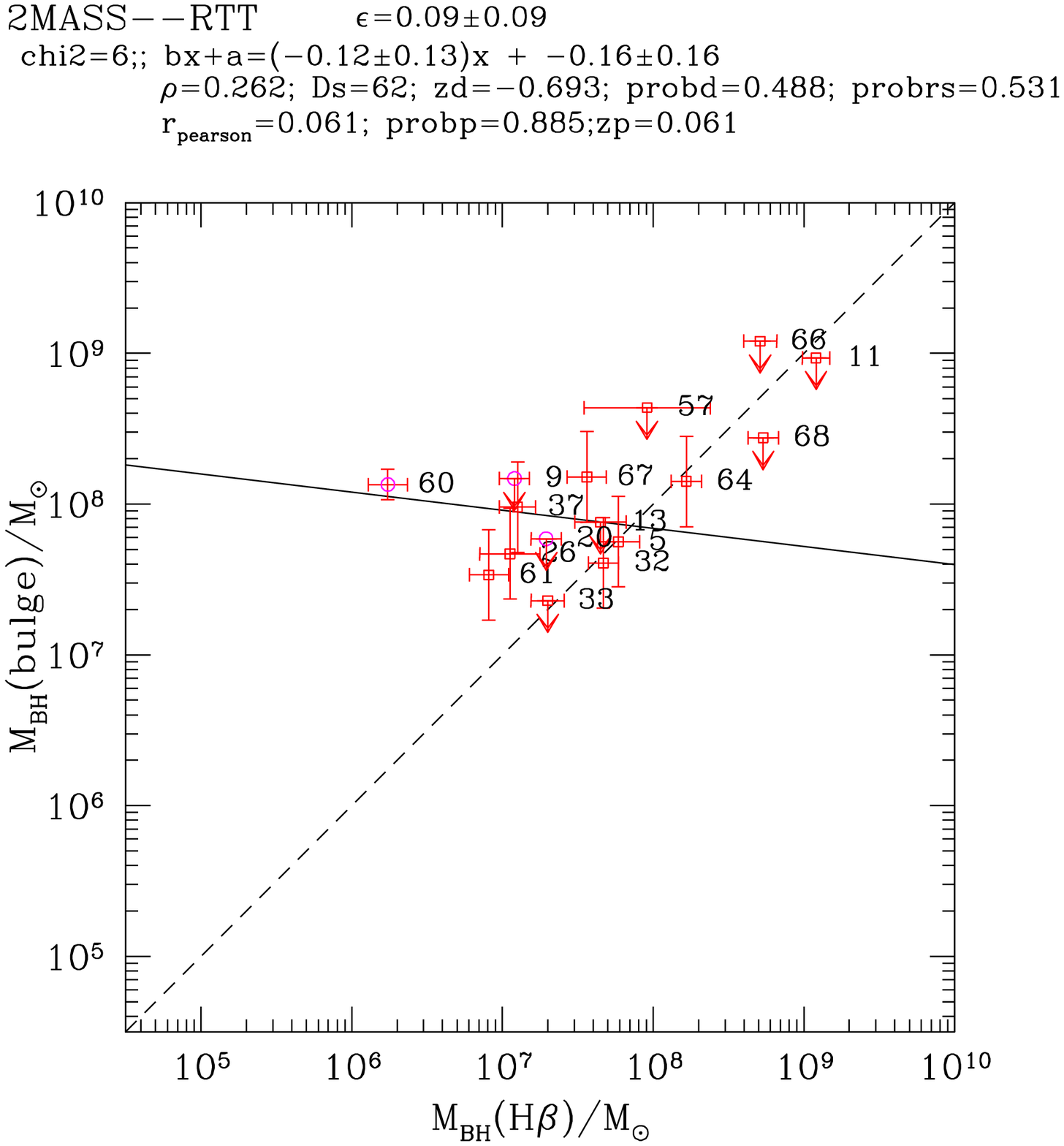}
}
\subfigure{
\includegraphics[width=0.8\columnwidth,viewport= 0 0 500 500,clip]{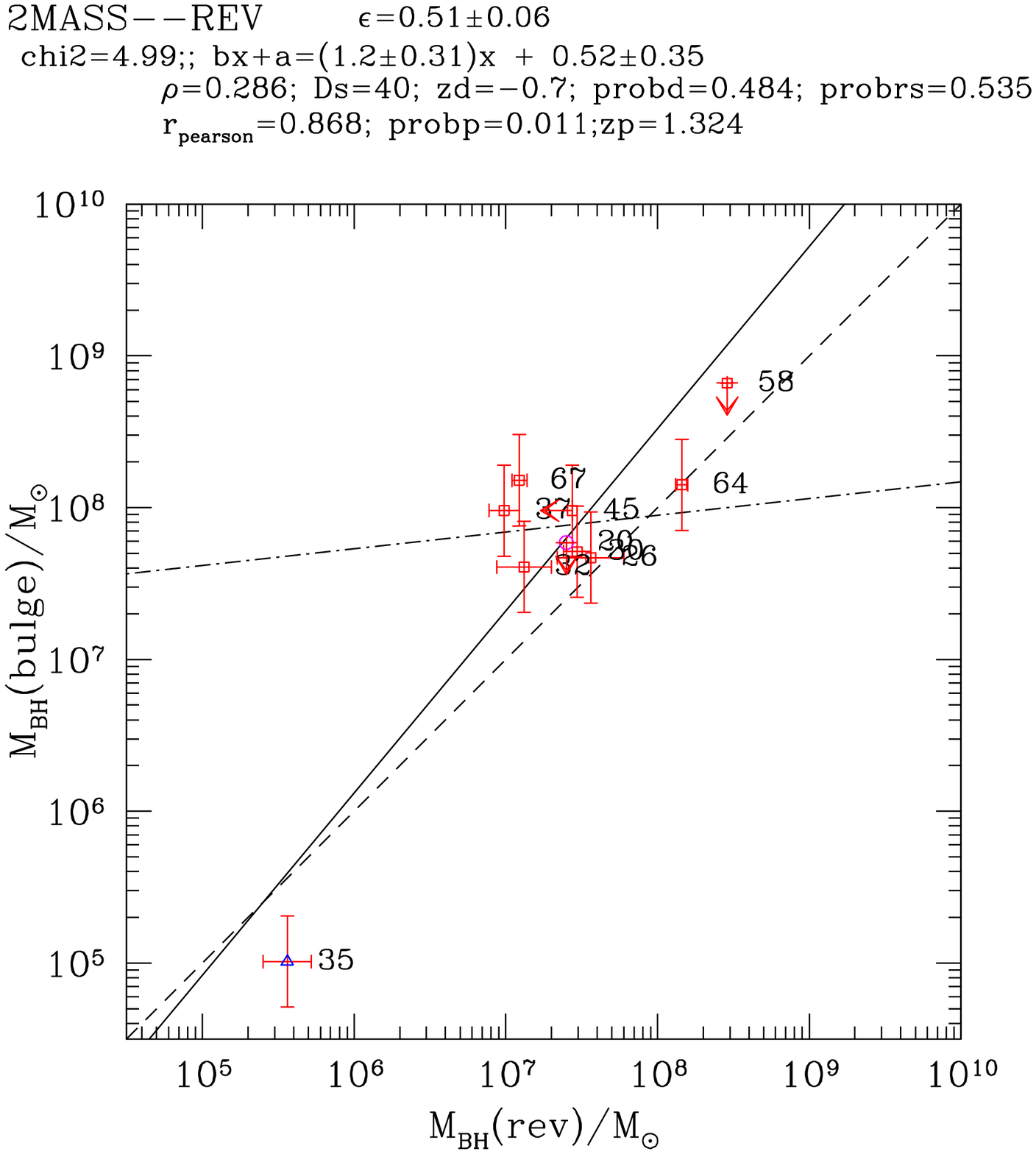}
}
\subfigure{
\includegraphics[width=0.8\columnwidth,viewport= 0 0 500 500,clip]{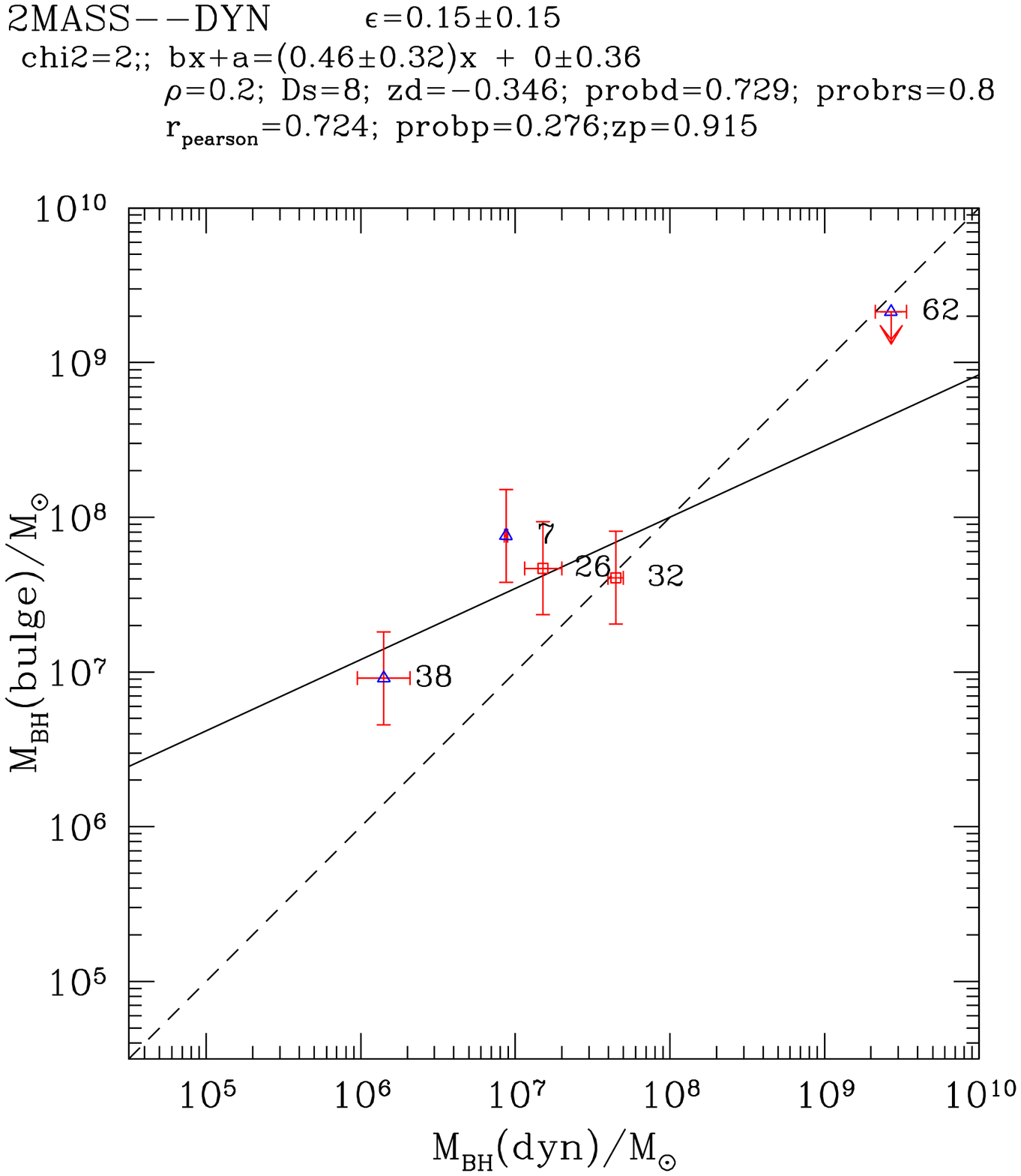}
}
\caption{Comparison of various SMBH mass estimates.
The regressions in logarithmic space are indicated by the
solid lines. In the middle panel, the dash-dotted line indicates
the regression withoutNGC 4395 (object \No 35). The dashed line indicates the ratio
1:1. TheSeyfert 1, 2,
and narrow-line Seyfert 1 (NLS1) galaxies are marked by
the squares, triangles, and circles, respectively. The ordinal
numbers of the objects are also given in the tables.}
\label{fig:corr}
\end{figure}

\begin{figure}[h!]
\centering
\SetFigLayout[2]{1}{2}
\subfigure{
\includegraphics[width=0.8\columnwidth,viewport= 0 0 500 500,clip]{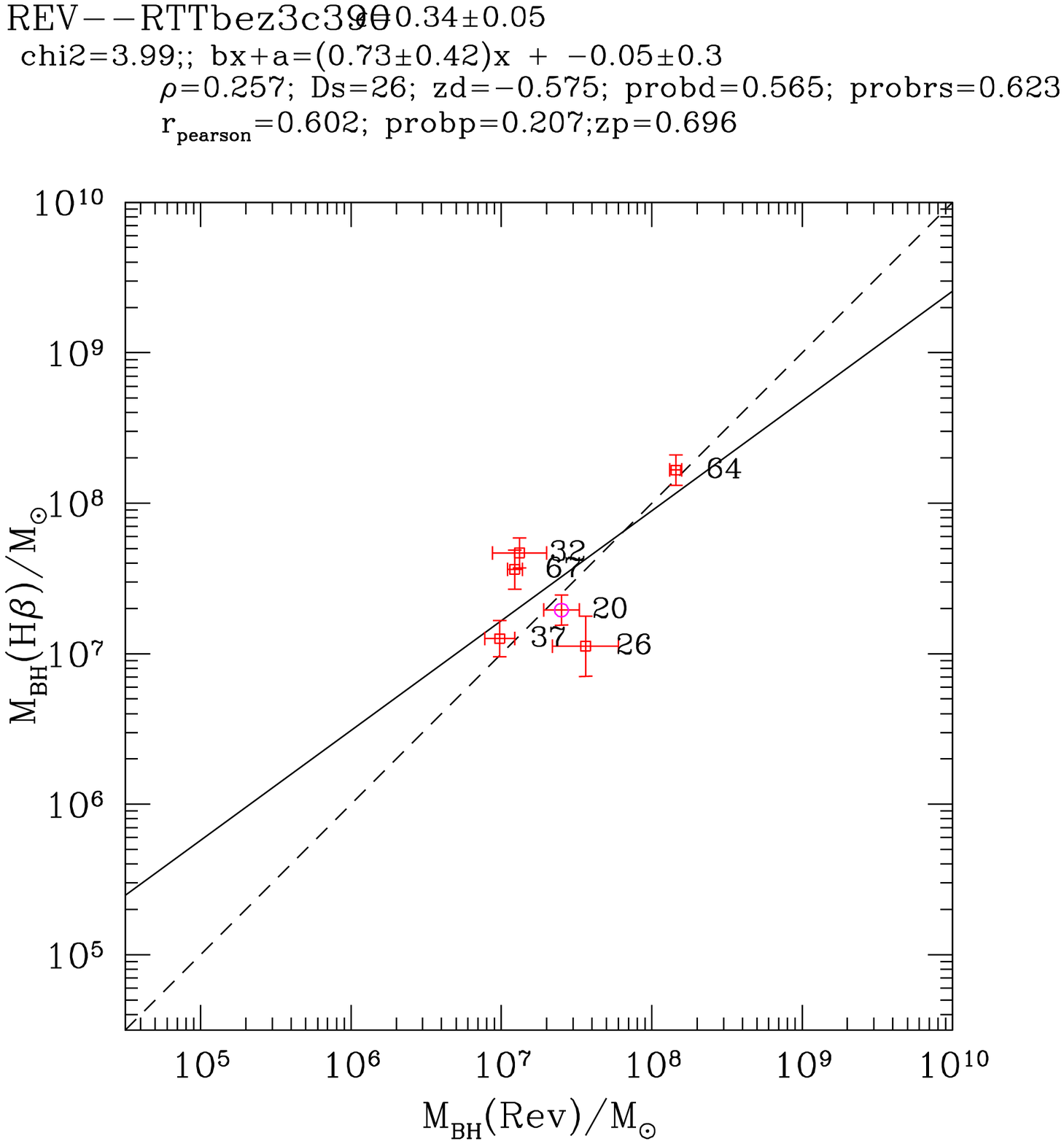}
}
\hfill
\subfigure{
\includegraphics[width=0.8\columnwidth,viewport= 0 0 500 500,clip]{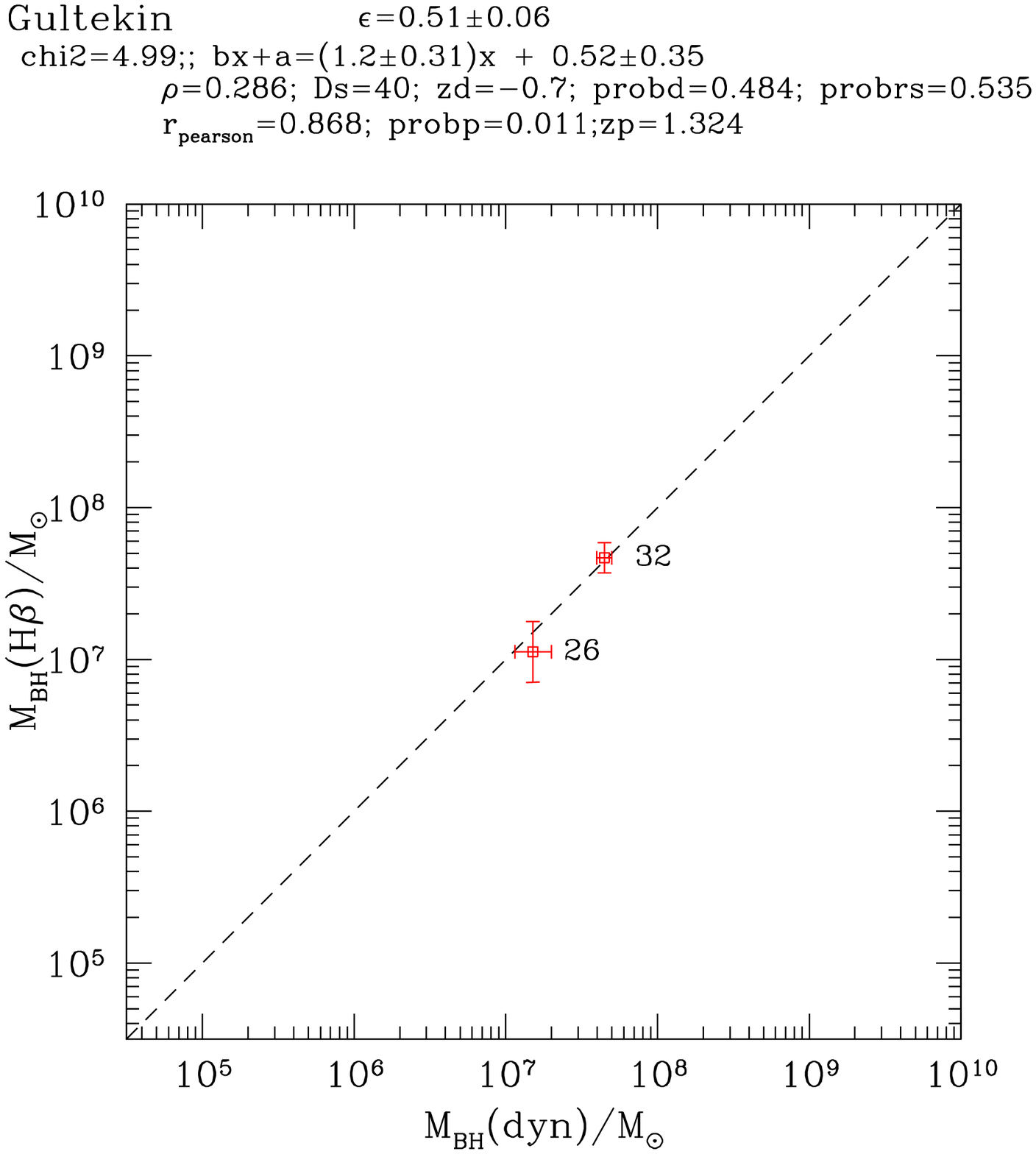}
}
\caption{Same as Fig.~\ref{fig:corr} for other correlations}
\label{fig:corr_hb}
\end{figure}

\begin{table*}
\centering
\caption{Cross-correlation of the SMBH masses found by different methods
\label{tab:regression}}
\smallskip

\scriptsize

\begin{tabular}{lcr@{$\pm$}lr@{$\pm$}lr@{$\pm$}lrccc}
\hline
\hline

\multicolumn{1}{c}{\mbox{Correlation}} &
\multicolumn{1}{c}{N} &
\multicolumn{2}{c}{$\beta \pm \Delta \beta$} &
\multicolumn{2}{c}{$\alpha \pm \Delta \alpha$} &
\multicolumn{2}{c}{$\epsilon_0 \pm \Delta \epsilon_0$} &
\multicolumn{1}{c}{$\rho$} &
\multicolumn{1}{c}{$P_\rho$} &
\multicolumn{1}{c}{$r$} &
\multicolumn{1}{c}{$P_r$} 
\\
\hline
$\Mhb$--$\Mbulge$ & 8 & -0.12 & 0.13 &-0.16 & 0.16 & \multicolumn{2}{c}{$\sim 0$} &0.26&0.53&0.06&0.89\\
$\Mrev$--$\Mbulge$ & 7 & 1.20 & 0.31 &0.52 & 0.35& 0.51 & 0.06 &0.29&0.54&0.87&0.01\\
$\Mrev$--$\Mbulge \mbox{(without NGC~4395)}$ & 6 & 0.11 & 0.30 &-0.05 & 0.23 & \multicolumn{2}{c}{$\sim 0$}&-0.14&0.79&0.18&0.73\\
$\Mdyn$--$\Mbulge$ & 4 & 0.46 & 0.32 & 0.01 & 0.35 &\multicolumn{2}{c}{$\sim 0$} &0.20 &0.80 &0.72&0.28\\
$\Mrev$--$\Mhb$ & 6 & 0.73 & 0.42 & -0.05 & 0.30 & 0.34 & 0.05 & 0.26 & 0.62 & 0.60 & 0.21\\
\hline
\end{tabular}
\begin{flushleft}
Note. $N$ is the number of objects; $\rho$ is the Spearman rank correlation coefficient; $P_\rho$ is the probability that the null hypothesis holds; $r$ is the Pearson correlation coefficient; $P_r$ is the probability that the null hypothesis holds.
\end{flushleft}
\end{table*}

\section{COMPARISON OF VARIOUS MASS
DETERMINATION METHODS}
\label{s:comp}
Table~\ref{tab:agn} gathers together the various estimates
of the SMBH masses in 68 AGNs from the INTEGRAL
survey. For 67 objects, we present the
estimates ($\Mbulge$) based on the infrared luminosity
of the bulge/elliptical galaxy obtained from 2MASS
data here. For 17 Seyfert 1 galaxies, we present
the estimates ($\Mhb$) based on the parameters of the
broad H$_\beta$ line that we obtained from RTT-150 data.
In addition, for ten Seyfert 1 galaxies, we provide the
reverberation-based measurements ($\Mrev$) \citep{petetal04,onkpet02,petetal05,denetal06}.
Finally, for five objects, we
give the masses ($\Mdyn$) measured from the kinematics
of the stars or gas near the SMBH \citep{guletal09}.
\par
The various estimates of the SMBH masses in the
AGNs of our sample are compared in Figs.~\ref{fig:corr} and \ref{fig:corr_hb}.
To get a quantitative idea of how the various methods
agree between themselves, we calculated the linear
regression:
\begin{eqnarray}
y=\alpha+\beta x\,,\ \
x=\lg\frac{M_1}{10^8 M_\odot}\,,\ \
y=\lg \frac{M_2}{10^8 M_\odot}\,,
\label{eq:regres}
\end{eqnarray}
where $M_1$ and $M_2$ are the SMBH masses estimated
by two different methods. Normalizing the masses
to $10^8 M_\odot$ (a typical value for our sample) allows the
correlation between the regression parameters $\alpha$ and
$\beta$ to be minimized.
\par
In our calculations, we used the technique described
in \citealt{treetal02} and the FITEXY
algorithm \citep{preetal92}. More specifically, the
regression parameters $\alpha$ and $\beta$ are found by minimizing
the function $\chi^2$:
\begin{equation}
\chi^2=\sum_{i=1}^N\dfrac{(y_i-\alpha-\beta x_i)^2}{\epsilon^2_{yi}+\epsilon^2_0+\beta^2\epsilon^2_{xi}} \,,	
\label{eq:chi2}
\end{equation}
where $x_{i}$ and $y_{i}$ are the estimates for the $i$th object,
$\epsilon_{xi}$ and $\epsilon_{yi}$ are the corresponding errors, and $\epsilon_{0}$ is the
intrinsic scatter of the correlation. The value of $\epsilon_{0}$ is
determined when $\chi^2$ divided by the number of degrees
of freedom becomes equal to one. The errors of $\alpha$, $\beta$,
and $\epsilon_{0}$ are found by varying the parameters within the
range defined by the condition $\chi^2-\chi^2_{\text{min}}\leq 1$
\par
Table~\ref{tab:regression} presents the results of our cross-correlation
analysis of the SMBH mass estimate from the bulge
luminosity ($\Mbulge$), which was used as the variable $y$
in Eq. (\ref{eq:regres}), with other types of estimates ($\Mhb$, $\Mrev$,
$\Mdyn$). The result of a similar comparison of $\Mrev$
with $\Mhb$ is also presented. In our calculations, we
disregarded the upper limits on the mass available for
several objects.
\par
We found no statistically significant correlation
between $\Mrev$ and $\Mhb$. This is not surprising, because
only six objects were compared. However,
this comparison confirms the validity of the empirical
formula (\ref{eq:vespet}) for Seyfert galaxies, because no systematic
bias of the $\Mhb$ estimates relative to the $\Mrev$
estimates was revealed, while the scatter of individual
values about the mean ($\epsilon_0=0.34\pm 0.05$) in our
minisample agrees well with that found by \citep{vespet06} when calibrating Eq.(\ref{eq:vespet})
based on reverberation observations of a larger sample
(28 objects) of Seyfert galaxies and quasars.
\par
As regards the determination of the SMBH mass
from the bulge luminosity, as can be seen from Fig.~\ref{fig:corr},
these estimates turn out to be systematically higher
than those obtained by other methods for $\Mbh\lesssim 10^{8} M_\odot$
(the number of objects for comparison is
too small for the corresponding conclusion to be formulated
with regard to higher masses). Our regression
analysis revealed no statistically significant
correlation of $\Mbulge$ with $\Mhb$ and $\Mdyn$. A weakly
significant correlation was found between $\Mrev$ and
$\Mbulge$, with the dependence found being in agreement
with a linear one ($\beta=1.2\pm 0.3$). However, only
one object, NGC 4395, makes a major contribution
to this correlation. It is often called a "dwarf Seyfert
galaxy" (see, e.g., \citealt{petetal05}) because of an
atypically low SMBH mass ($\Mbh<10^6 M_\odot$) and a
low luminosity (its luminosity ismuch lower than that
of the remaining objects in our sample, see Table~\ref{tab:agnintro}).
If we exclude NGC 4395 from consideration, then the
correlation of $\Mrev$ with $\Mbulge$ (based on six objects)
ceases to be statistically significant.
\par
The main conclusion that can be drawn from this
study is that the masses $\Mbh$ estimated from the
infrared bulge luminosity are in poor agreement and,
on average, are overestimated relative to the more
reliable estimates of the SMBH masses in Seyfert
galaxies. The differences of the correlation between
the SMBH mass and the bulge luminosity for AGNs
and normal galaxies have already been pointed out
in several previous papers \citep{neletal04,benetal09}.

\begin{table*}

\caption{Summary table of various SMBH mass estimates
\label{tab:agn} }

\smallskip

\scriptsize

\begin{tabular}{lccccD{.}{.}{2}D{.}{.}{2}cD{.}{.}{2}D{.}{.}{2}D{.}{.}{2}}
\hline
\hline
\multicolumn{1}{c}{Object} &
\multicolumn{1}{c}{\No} &
\multicolumn{1}{c}{$\lg\Mdyn$} &
\multicolumn{1}{c}{$\lg\Mrev$} &
\multicolumn{1}{c}{$\lg\Mhb$} &
\multicolumn{1}{c}{$\lg\Mbulge$} &
\multicolumn{1}{c}{$\Leddbulge$} &
\multicolumn{1}{c}{$\Lbol$} &
\multicolumn{1}{c}{$\Lbol/$} &
\multicolumn{1}{c}{$\Leddbest$} &
\multicolumn{1}{c}{$\Lbol/$}
\\   
\multicolumn{1}{c}{} &  
\multicolumn{1}{c}{} &   
\multicolumn{1}{c}{($M_\odot$)} & 
\multicolumn{1}{c}{($M_\odot$)} &
\multicolumn{1}{c}{($M_\odot$)} & 
\multicolumn{1}{c}{($M_\odot$)} &
\multicolumn{1}{c}{(erg/s)} &
\multicolumn{1}{c}{(erg/s)} &     
\multicolumn{1}{c}{$\Leddbulge$} &
\multicolumn{1}{c}{(erg/s)} &
\multicolumn{1}{c}{$\Leddbest$}
\\
\hline
MRK348&1&&&&<7.71&<45.81&44.51&>0.051&<45.81&>0.051\\ 
MCG-01-05-047&2&&&&7.58&45.68&43.97&0.020&45.68&0.020\\ 
NGC788&3&&&&7.92&46.02&44.23&0.016&46.02&0.016\\ 
LEDA138501&4&&&$ 8.53_{8.45}^{8.61} $&&&45.29&&46.63&0.046\\ 
MRK1040&5&&&$ 7.77_{7.63}^{7.91} $&7.75&45.85&44.42&0.038&45.87&0.036\\ 
IGRJ02343+3229&6&&&&8.13&46.23&44.29&0.012&46.23&0.012\\ 
NGC1068&7&$ 6.94_{6.92}^{6.95} $&&&7.88&45.98&44.57&0.039&45.04&0.339\\ 
NGC1142&8&&&&8.48&46.58&44.88&0.020&46.58&0.020\\ 
1H0323+342&9&&&$ 7.08_{6.98}^{7.18} $&<8.17&<46.27&45.32&>0.114&45.18&1.398\\ 
NGC1365&10&&&&7.97&46.07&43.06&0.001&46.07&0.001\\ 
3C111&11&&&$ 9.08_{8.99}^{9.17} $&<8.97&<47.07&45.57&>0.032&47.18&0.025\\ 
ESO033-G002&12&&&&7.82&45.92&44.09&0.015&45.92&0.015\\ 
IRAS05078+1626&13&&&$ 7.65_{7.48}^{7.82} $&\lesssim7.88& \lesssim 45.98&44.56& \gtrsim 0.039&45.75&0.065\\ 
ESO005-G004&14&&&&7.50&45.60&43.13&0.003&45.60&0.003\\ 
MRK3&15&&&&7.96&46.06&44.38&0.021&46.06&0.021\\ 
MRK6&16&&&
&8.01&46.11&44.40&0.020&46.11&0.020\\ 
IGRJ07563-4137&17&&&&8.02&46.12&44.03&0.008&46.12&0.008\\ 
ESO209-G012&18&&&&8.49&46.59&44.74&0.014&46.59&0.014\\ 
IRAS09149-6206&19&&&&\backsimeq9.59& \backsimeq 47.69&45.14&\backsimeq 0.003&\backsimeq 47.69&\backsimeq 0.003\\ 
MRK110&20&&$ 7.40_{7.28}^{7.49} $(1)&$ 7.29_{7.19}^{7.39} $&<7.77&<45.87&45.16&>0.197&45.50&0.463\\ 
IGRJ09446-2636&21&&&&\lesssim8.62& \lesssim 46.72&46.26&\gtrsim 0.351&\lesssim 46.72&\gtrsim 0.351\\ 
NGC2992&22&&&&7.62&45.72&43.71&0.010&45.72&0.010\\ 
MCG-5-23-16&23&&&&7.28&45.38&44.13&0.057&45.38&0.057\\ 
NGC3081&24&&&&7.55&45.65&43.72&0.012&45.65&0.012\\ 
ESO263-G013&25&&&&7.98&46.08&44.67&0.039&46.08&0.039\\ 
NGC3227&26&$ 7.18_{6.85}^{7.30} $&7.59(2)&$ 7.05_{6.85}^{7.25} $&7.67&45.77&43.62&0.007&45.28&0.022\\ 
NGC3281&27&&&&8.07&46.17&43.93&0.006&46.17&0.006\\ 
IGRJ10386-4947&28&&&&\lesssim8.27& \lesssim 46.37&45.04&\gtrsim 0.047&\lesssim 46.37&\gtrsim 0.047\\ 
IGRJ10404-4625&29&&&&7.88&45.98&44.37&0.025&45.98&0.025\\ 
NGC3783&30&&$ 7.47_{7.38}^{7.54} $(1)&&7.71&45.81&44.29&0.031&45.57&0.053\\ 
IGRJ12026-5349&31&&&&8.20&46.30&44.57&0.019&46.30&0.019\\ 
NGC4151&32&$ 7.65_{7.6}^{7.7} $&$ 7.12_{6.94}^{7.25} $(1)&$ 7.67_{7.57}^{7.77} $&7.61&45.71&44.32&0.041&45.75&0.038\\ 
MRK50&33&&&$ 7.30_{7.19}^{7.41} $&<7.36&<45.46&44.14&>0.048&45.40&0.056\\ 
NGC4388&34&&&&7.20&45.30&43.74&0.028&45.30&0.028\\ 
NGC4395&35&&$ 5.56_{5.40}^{5.67} $(3)&&5.01&43.11&41.55&0.028&43.66&0.008\\ 
NGC4507&36&&&&7.81&45.91&44.46&0.036&45.91&0.036\\ 
NGC4593&37&&$ 6.99_{6.89}^{7.08} $(4)&$ 7.10_{6.98}^{7.22} $&7.98&46.08&43.99&0.008&45.09&0.080\\ 
NGC4945&38&$ 6.15_{5.95}^{6.32} $&&&6.96&45.06&42.49&0.003&44.25&0.018\\ 
ESO323-G077&39&&&&8.16&46.26&44.08&0.007&46.26&0.007\\ 
IGRJ13091+1137&40&&&&7.91&46.01&44.63&0.042&46.01&0.042\\ 
IGRJ13149+4422&41&&&&7.95&46.05&44.76&0.052&46.05&0.052\\ 
CENA&42&&&&8.24&46.34&42.89&0.0004&46.34&0.0004\\ 
MCG-6-30-15&43&&&&7.26&45.36&43.61&0.018&45.36&0.018\\ 
MRK268&44&&&&7.95&46.05&44.74&0.050&46.05&0.050\\ 
IC4329A&45&&$<7.40_{}^{} $(1)&&7.98&46.08&44.90&0.067&<45.50&>0.254\\ 
NGC5506&46&&&&7.68&45.78&44.07&0.020&45.78&0.020\\ 
IGRJ14552-5133&47&&&&7.23&45.33&43.84&0.033&45.33&0.033\\ 
IGRJ14561-3738&48&&&&8.25&46.35&44.22&0.008&46.35&0.008\\ 
IC4518A&49&&&&\lesssim7.71& \lesssim 45.81&44.06& \gtrsim 0.018& \lesssim 45.81& \gtrsim 0.018\\ 
WKK6092&50&&&&7.54&45.64&43.89&0.018&45.64&0.018\\ 
IGRJ16185-5928&51&&&&7.39&45.49&44.62&0.137&45.49&0.137\\ 
ESO137-G34&52&&&&8.02&46.12&43.43&0.002&46.12&0.002\\ 
IGRJ16482-3036&53&&&&\lesssim7.89& \lesssim 45.99&44.70& \gtrsim 0.052& \lesssim 45.99& \gtrsim 0.052\\ 
NGC6221&54&&&&7.41&45.51&42.89&0.002&45.51&0.002\\ 
IGRJ16558-5203&55&&&&\lesssim9.07& \lesssim 47.17&45.24&\gtrsim 0.012&\lesssim 47.17& \gtrsim 0.012\\ 
NGC6300&56&&&&7.59&45.69&43.02&0.002&45.69&0.002\\ 
IGRJ17418-1212&57&&&$ 7.96_{7.54}^{8.38} $&\lesssim8.64& \lesssim 46.74&44.85& \gtrsim 0.013&46.06&0.062\\ 
3C390.3&58&&$ 8.46_{8.35}^{8.55} $(1)&$ 9.61_{9.43}^{9.79} $&<8.82&<46.92&45.60&>0.048&46.56&0.111\\ 
IGRJ18559+1535&59&&&&\lesssim8.35& \lesssim 46.45&45.53& \gtrsim 0.122&\lesssim 46.45& \gtrsim 0.122\\ 
1H1934-063&60&&&$ 6.24_{6.11}^{6.37} $&8.13&46.23&43.58&0.002&44.34&0.176\\ 
NGC6814&61&&&$ 6.91_{6.78}^{7.04} $&7.53&45.63&43.42&0.006&45.01&0.026\\ 
CYGA&62&$ 9.43_{9.28}^{9.53} $&&&<9.33&<47.43&45.57&>0.014&47.53&0.011\\ 
IGRJ20286+2544&63&&&&7.12&45.22&44.11&0.079&45.22&0.079\\ 
MRK509&64&&$ 8.16_{8.12}^{8.20} $(1)&$ 8.22_{8.12}^{8.32} $&8.15&46.25&45.11&0.073&46.26&0.072\\ 
NGC7172&65&&&&7.81&45.91&43.87&0.009&45.91&0.009\\ 
MR2251-178&66&&&$ 8.71_{8.6}^{8.82} $&<9.08&<47.18&45.60&>0.027&46.81&0.062\\ 
NGC7469&67&&$ 7.09_{7.04}^{7.14} $(1)&$ 7.56_{7.43}^{7.69} $&8.18&46.28&44.38&0.013&45.19&0.157\\ 
MRK926&68&&&$ 8.73_{8.63}^{8.83} $&<8.44&<46.54&45.20&>0.046&46.83&0.024\\ 

\hline
\end{tabular}
Note. $\Mdyn$ is the dynamical mass estimate \citep{guletal09}; $\Mrev$ is the reverberation-based mass estimate from: (1) \cite{petetal04}, (2) \cite{onkpet02}, (3) \cite{petetal05}, (4)
\cite{denetal06}; $\Mhb$ is the mass estimate from the parameters of the broad H$_\beta$ line (RTT-150);
$\Mbulge$ is themass estimate from the bulge luminosity(2MASS); $\Lbol$ is the bolometric luminosity; $\Leddbulge$ is the Eddington luminosity calculated from $\Mbulge$; $\Leddbest$ is the Eddington luminosity calculated from the best mass estimate. 

\end{table*}

\section{THE ACCRETION RATE}
\label{s:accr}
Given the SMBH mass estimate, we can calculate
the Eddington limit,
\begin{equation}
\Ledd=1.3\times 10^{46}\frac{\Mbh}{10^8 M_\odot},
\label{eq:edd}
\end{equation} 
and compare it with the bolometric AGN luminosity.
In our recent paper \citep{sazetal12},
based on the set of Spitzer infrared observations and
INTEGRAL hard X-ray observations for the same
sample of objects that is investigated here, we showed
that the hard X-ray luminosity is a good indicator of
the bolometric AGN luminosity, more specifically,
\begin{equation} 
\frac{\Lbol}{\Lint}\sim 9.
\label{eq:bol}
\end{equation}
\par
Table~\ref{tab:agn} gives the values of $\Ledd$, $\Lbol$ and $\Lbol/\Ledd$
obtained for our objects from these formulas using
two alternative types of SMBH mass estimation:
(1) $\Mbulge$ (from the correlation with the infrared luminosity of the bulge/elliptical galaxy); 
(2) the best estimate~--- if there is a dynamical mass estimate
($\Mdyn$) for an object, then we use it; otherwise (in
order of preference), the estimate by reverberation
mapping ($\Mrev$), from the parameters of the broad H$_\beta$
line ($\Mhb$), and from the correlation with the bulge
luminosity ($\Mbulge$) is used.
\par
Formula (\ref{eq:bol}) cannot be used for the Compton-thick
Seyfert galaxy NGC 1068. Therefore, we estimated
its bolometric luminosity from the infrared luminosity
($\lg\Lmir=43.87$) of the dusty torus \citep{masetal06} based on the relation
\begin{equation}
\frac{\Lbol}{\Lmir}\sim 
5
\label{eq:bol_mir}
\end{equation}
derived by \citealt{sazetal12} from Spitzer observations
of Seyfert galaxies from the INTEGRAL survey.
\par
\begin{figure}[!]
\centering
\SetFigLayout{1}{2}
\subfigure{
\includegraphics[width=\columnwidth,viewport= 0 0 550 550,clip]{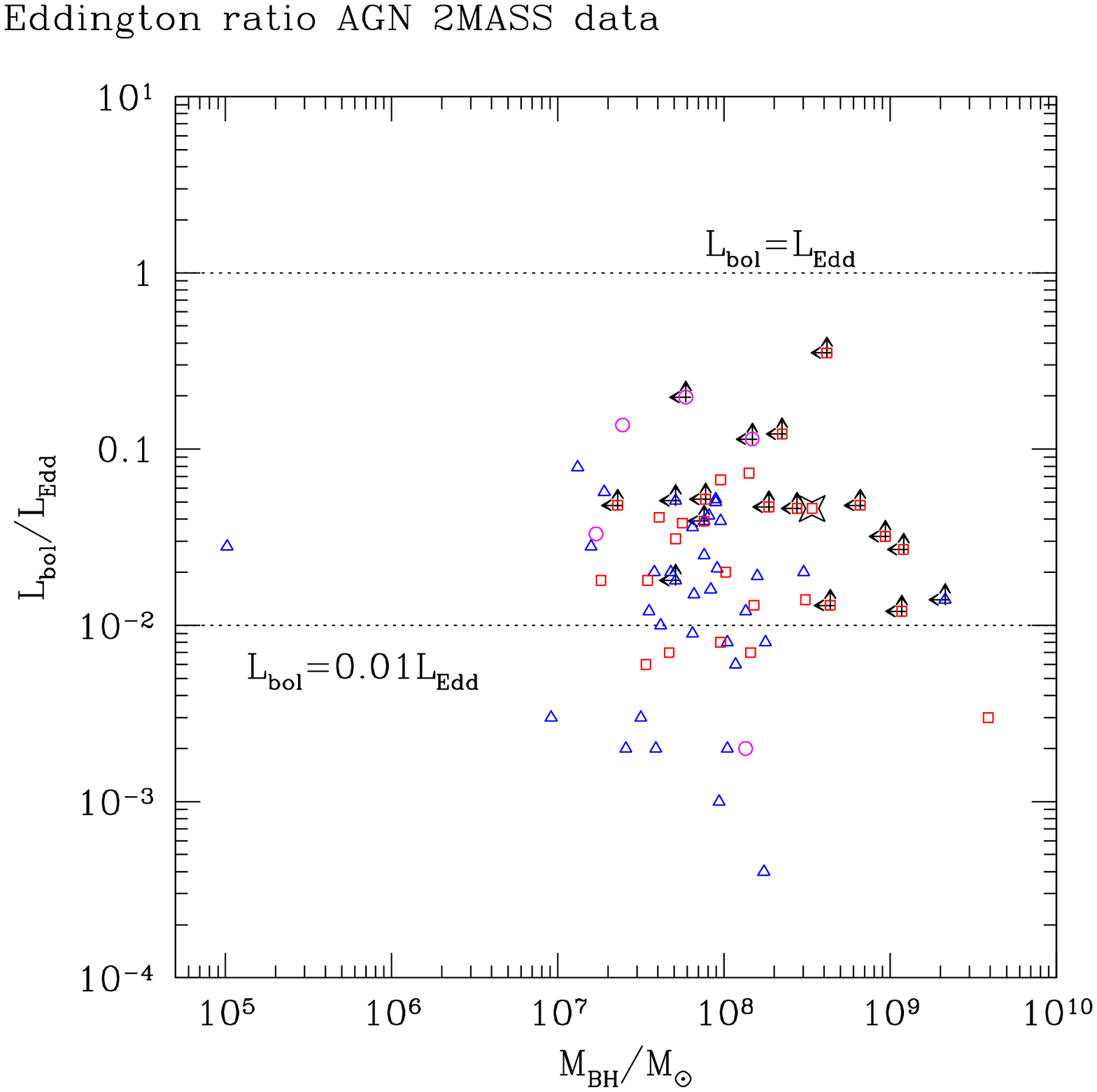}
}
\hfill
\subfigure{
\includegraphics[width=\columnwidth,viewport= 0 0 550 550,clip]{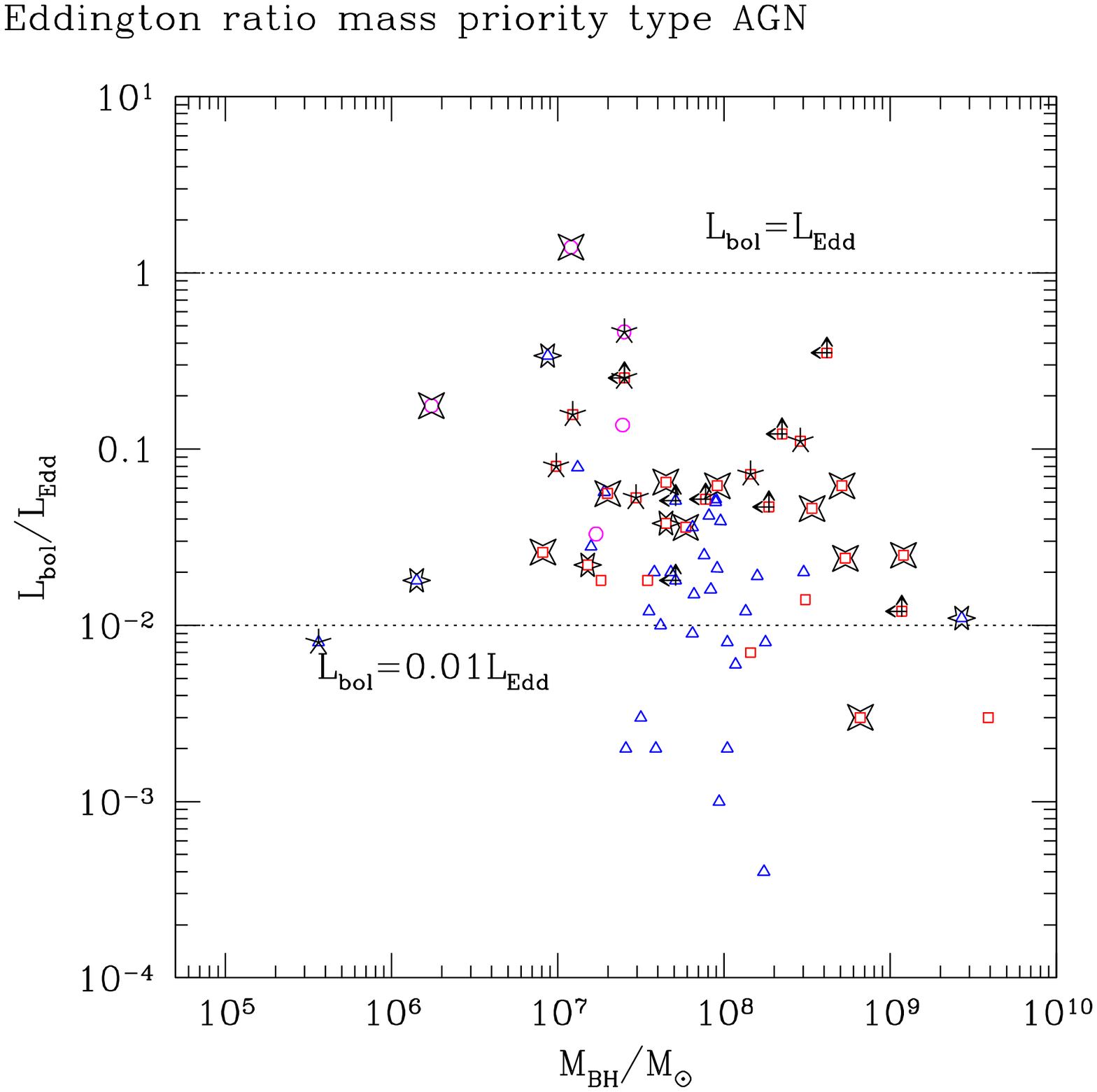}
}
\caption{{\sl Top:} Bolometric-to-Eddington AGN luminosity ratio versus SMBH mass. $\Mbh$ was estimated from the infrared bulge
luminosity, except for LEDA~138501 (marked by the four-point star) for which the estimate from the parameters of the broad H$_\beta$ line is presented. The Seyfert 1, 2, and narrow-line Seyfert 1 (NLS1) galaxies are marked by the squares, triangles, and
circles, respectively. The measurement uncertainties are not shown. {\sl Bottom:}
The same, only based on the best mass estimates: $\Mdyn$
(six-point stars), $\Mrev$ (five-point stars), $\Mhb$
(four-point stars), $\Mbulge$ (remaining points).}
\label{fig:lum}
\end{figure}

The ratio $\Lbol/\Ledd$ may be considered as an indicator
of the SMBH accretion regime. Figure~\ref{fig:lum}
presents the values of $\Lbol/\Ledd$ calculated both from
the Mbulge estimates (a) and from the best SMBH
mass estimates (b). For most of the objects, $0.01 < \Lbol/\Ledd <1$. This suggests that the gas is most
likely accreted onto the SMBH in a radiatively efficient regime via a geometrically thin, optically thick disk \citep{shasun73}.
\par
However, the bolometric luminosity for AGN
1H~0323+342, a narrow line Seyfert 1 (NLS1) galaxy,
exceeds the Eddington one. For two more NLS1 objects,
MRK~110 and 1H~1934-063, $\Lbol/\Ledd$ turns
out to be also high if the best SMBH mass estimate,
$\Mhb$, is used. In addition, $\Mbulge$ for 1H~1934-063
exceeds $\Mhb$ by two orders of magnitude. For the
remaining two NLS1 objects, IGR J14552-5133 and IGR J16185-5958, $\Lbol/\Ledd=0.03$ and 0.14,
respectively, but these estimates cannot be considered
reliable, because they were obtained from $\Mbulge$.
On the whole, these results confirm that the most
rapid SMBH growth occurs at the present epoch in
narrow-line Seyfert 1 galaxies \citep{mathur00} and
suggest that this growth is far from completion.
\par
For several objects, $\Lbol/\Ledd<0.01$ (Fig.~\ref{fig:lum}).
This may imply that the gas is accreted onto the
SMBH at a low rate and/or with a low electromagnetic wave emission efficiency. However, in all these
cases, the values of $\Lbol/\Ledd$ were obtained from
the $\Mbulge$ estimates and, therefore, may be grossly
underestimated.

\section{CONCLUSIONS}
\label{s:summary}
Our study based on a representative sample of
nearby AGNs has shown that the estimate of the
SMBH masses in Seyfert galaxies from the infrared
luminosity (and, consequently, from the mass) of the
stellar bulge cannot be considered reliable and, on
average, leads to overestimated values of $\Mbh$ (for
$\lesssim 10^8 M_\odot$). This may indicate that the black hole
growth in an appreciable fraction of Seyfert galaxies
is still far from completion; at least we can say
about this with confidence in regard to NLS1 Seyfert
galaxies. However, since here we used a rather small
sample of AGNs and since our estimates have considerable
uncertainties (in particular, we used rather
rough estimates of the bulge fraction in the galaxy's
total luminosity), this conclusion cannot be deemed
the final one. In addition, it is not yet clear how
it is reconciled with the conclusion reached by several
authors (see the Introduction) that AGNs obey
the standard $\Mbh$--$\sigstar$ correlation. The ratio of the
bolometric luminosity to the critical Eddington one
ranges from 1 to 100\% for the overwhelming majority
of objects. This suggests that the gas is accreted
onto the SMBHs in Seyfert galaxies at a high rate
and in a radiatively efficient regime. This conclusion
is consistent with the results of previous studies (see,
e.g., \cite{midetal08}, \cite{derosaetal12}).

\begin{acknowledgements}
\section{ACKNOWLEDGMENTS}
We wish to thank the referee for a number of useful
remarks. We used data from the Russian--Turkish
1.5-m telescope, the 2MASS survey (University of
Massachusetts, IPAC), and the NASA/IPAC/NED
(Jet Propulsion Laboratory, California Institute of
Technology), VizieR (CDS, Strasbourg), and Hyper-
Leda (Observatoire de Lyon) databases. This study
was supported by the Russian Foundation for Basic
Research (project nos. 09-02-00867-a, 10-02-
01442-a, 11-02-12285-ofi-m-2011, 11-02-12271-
ofi-m), Programs P-21 and OFN-16 of the Russian
Academy of Sciences, and the Program for Support of
Leading Scientific Schools of the Russian Federation
(NSh-5069.2010.2).
\end{acknowledgements}

Translated by V. Astakhov

\end{document}